\documentclass[notitlepage,aps,twocolumn,superscriptaddress,longbibliography]{revtex4-1}
\usepackage{amsmath}
\usepackage{amssymb}
\usepackage{physics}
\usepackage{todonotes}
\usepackage[unicode=true,colorlinks=true,citecolor=blue,urlcolor=blue]{hyperref}
\usepackage{footnote}
\usepackage{bm}
\usepackage{epsfig}
\usepackage{lipsum}
\usepackage{mathtools}
\usepackage{float}

\newcommand{\intdx}{\int \mathrm{d} x \,}
\newcommand{\crr}{\hat{c}_R }
\newcommand{\cll}{\hat{c}_L}
\newcommand{\ddx}[1]{\frac{\partial #1}{\partial x}}

\begin{document}
\title{
Detecting the relative phase between different frequency components of a photon using a three-level $\Lambda$ atom coupled to a waveguide}

\author{Janet Zhong}
\affiliation{Department of Applied Physics, Stanford University, Stanford, California 94305, USA}
\author{Rituraj}
\affiliation{Department of Electrical Engineering, Indian Institute of Technology Kanpur, Kanpur 208016, India}
\author{Fatih Dinc}
\affiliation{Department of Applied Physics, Stanford University, Stanford, California 94305, USA}
\author{Shanhui Fan}
\email{shanhui@stanford.edu}
\affiliation{Department of Applied Physics, Stanford University, Stanford, California 94305, USA}
\affiliation{Department of Electrical Engineering, Ginzton Laboratory, Stanford University, Stanford, California 94305, USA}

\begin{abstract}
We study the scattering inside
a waveguide of a single photon with a single three-level $\Lambda$ atom, both of
which are in a superposition state. The photon is in a superposition
state of two frequencies, whereas the atoms is in a superposition
state of two non-degenerate ground states. We find that the scattering depends on both the relative phase between the photon frequencies and the relative phase between the atomic ground states. Our results show that a three-level $\Lambda$ atom coupled to a waveguide can be used as photon phase filter that could detect the relative phase between the two frequencies of the photon superposition state. 
\end{abstract}
\date{\today}

\maketitle

\section{Introduction}
There have been many studies on the generation~\cite{zakka2011quantum,clemmen2016ramsey,aguayo2022integrated} or manipulation~\cite{tanzilli2005photonic,karnieli2018frequency,lukens2017frequency,lu2020fully,lu2019controlled,lu2018quantum,olisager2010frequency} of a single photon in a superposition of two frequencies (or `colours'). These frequency-superposition states have been called `color qubits'~\cite{aguayo2022integrated}, bichromatic photons~\cite{clemmen2016ramsey}, or frequency-bin qubits~\cite{lukens2017frequency}. Such states can be generated in the process of quantum frequency conversion, where a photon's frequency is converted while its quantum information is maintained, as experimentally demonstrated three decades ago~\cite{huang1992observation}. These states also arise in quantum frequency mixing (where different superposition states are converted to and from each other) which has been demonstrated using $\chi^{(2)}$ nonlinearities~\cite{kobayashi2016frequency,kobayashi2017coherent}, $\chi^{(3)}$ nonlinearities~\cite{mcguiness2010quantum,clemmen2016ramsey,dmitriev2017quantum} as well as electro-optic modulation~\cite{lu2018quantum,lu2020fully}.

The manipulation of frequency superposition states at the single-photon level has generated interest for its potential applications in quantum information processing~\cite{lukens2017frequency}. Building on the work in linear optical quantum computation~\cite{knill2001scheme}, frequency-bin qubits have been proposed as an alternative to other photonic qubits such as those based on polarization. Polarization-based qubits can only have a two-dimensional Hilbert space, whereas frequency qubits can have a Hilbert space with much higher dimension. Frequency qubits may also be scalable for on-chip photonics~\cite{ultrabroadband2021javid}. In recent years, it has been theoretically proposed and experimentally verified that it is possible to achieve arbitrary single-qubit gates in frequency space~\cite{lu2020fully,lu2019controlled}. Aside from quantum information processing applications, frequency superposition states may also be useful for quantum metrology~\cite{giovannetti2011advances} or phase-spectroscopy~\cite{rinehart2012quantitative} at the low-light level. 
\begin{figure}[h]
\centering
\includegraphics[width=0.45\textwidth]{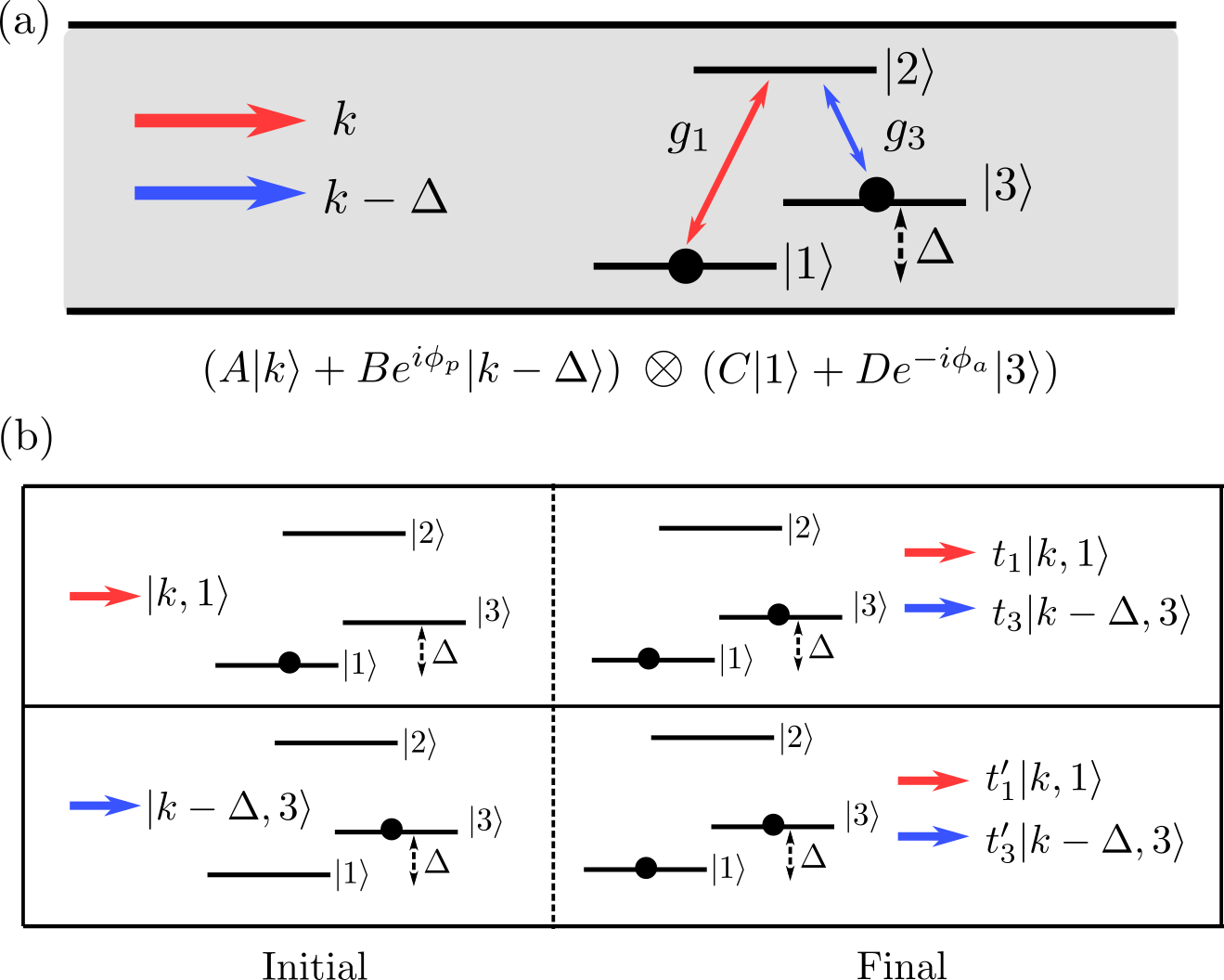}
\caption{(a) A $\Lambda$-type 3-level interacting with a photon propagating in single-mode waveguide (highlighted as gray). The photon is in frequency-superposition state. The photon frequencies $k,k-\Delta$ are in resonance with the atomic transition frequencies $E_2-E_1$ and $E_2-E_3$. (b) Transmission and reflection pathways for atom initially in $\ket{k,1}$ and $\ket{k-\Delta,3}$, where $k$ is kept general. Note that both scenarios have two scattering output channels, in which the atoms are in either the $\ket{1}$ or $\ket{3}$ state. Here $t_1, r_1,t_3, r_3$ is the transmission and reflection of the scattering where the atomic final state is the $\ket{1}$ and $\ket{3}$ state respectively and the initial atomic state is $\ket{1}$. Similarly, $t_1^{\prime}, r_1^{\prime},t_3^{\prime}, r_3^{\prime}$ is the transmission and reflection where the atomic final state is the $\ket{1}$ and $\ket{3}$ state respectively and the initial atomic state is $\ket{3}$. }
\label{maysetup}
\end{figure}

Although the generation and manipulation of single-photon frequency superposition states has been well-studied~\cite{yesharim2022observation,zakka2011quantum,clemmen2016ramsey,aguayo2022integrated,tanzilli2005photonic,karnieli2018frequency,lukens2017frequency,lu2020fully,lu2019controlled,lu2018quantum,olisager2010frequency}, there is less attention devoted to the measurement of the relative phase between different frequency components of such states. Most existing works focus on measuring the amplitude spectrum of the state, where the phase information is lost. The relative phase of two frequency components of a single photon was measured in Ref.~\cite{clemmen2016ramsey} using single-photon Ramsey interference. Here, they generated arbitrary frequency-superposition states using Bragg scattering four-wave mixing which is a $\chi^{(3)}$ nonlinear process. The relative phase between different frequency components could be measured because the outcome of their projective measurement depends on this relative phase. A related study uses electro-optic modulators instead of nonlinear processes~\cite{lu2020fully}. Their set-up also allows them to achieve arbitrary frequency qubit rotations on the Bloch sphere and the output state can then be measured using quantum state tomography, where the relative phase between frequency components can be determined using Pauli measurements.

In this study, we propose a novel way to measure the relative phase of different frequency components of a single photon using a setup in waveguide quantum electrodynamics~\cite{sheremet2021waveguide,roy2017colloquium,turschmann2019coherent,chang2018colloquium}. Our proposed set-up consists of a three-level atom coupled to a one-dimensional waveguide as seen in Fig.~\ref{maysetup}(a). The atom is of the $\Lambda$ type with a two-dimensional ground state manifold spanned by eigenstates $\ket{1}$ and $\ket{3}$ at energy $E_1$ and $E_3$, respectively, and an excited state $\ket{2}$ at energy $E_2$.   It has been shown that one can prepare the atom to be in an arbitrary ground state in this two-dimensional manifold~\cite{djotyan2008multiphoton,he2009arbitrary,randall2018generation}. Here we scatter a single-photon with two frequency components near $E_2-E_1$ and $E_2-E_3$ (where $\hbar=1$) against the atom. We show that with a suitable choice of the atomic ground state, the outcome of the scattering process can be used to determine the relative phase of the two photon frequency components. 
Scattering for a $\Lambda$ atom in a superposition of ground states was studied in Ref.~\cite{martens2013photon}. Quantum state transfer~\cite{chan2022quantum} and quantum frequency conversion~\cite{bradford2012single} has also been studied in a $\Lambda$-type atom. However, none of these studies specifically study the dependence of the relative photon phase on the transport for a $\Lambda$ atom coupled to a waveguide, which is the focus of this work. While the measurement of the relative photon phase of two frequency superposition states was done in EOM~\cite{lu2020fully} and Ramsey interference setups~\cite{clemmen2016ramsey}, the benefit of using our setup over these methods, is that our waveguide quantum electrodynamics system can compare the quantum information of two objects. That it is, the scattering output of our system can allow us to measure both the relative phase of the frequency superposition states as well as the relative phase of the atomic ground states in the same platform, as opposed to just one of these phases as in Ref.~\cite{lu2020fully, clemmen2016ramsey}. \\

\section{Model}

To treat the system of Fig.~\ref{maysetup}(a), we use a real space formalism originally developed in Ref.~\cite{shen2005coherent} and applied to three level atoms in Ref.~\cite{witthaut2010photon}. Let $g_1, g_3$ be the atom-photon coupling constants for the $\ket{1} \rightarrow \ket{2}$ and $\ket{3} \rightarrow \ket{2}$ transitions and assume the $\ket{1} \rightarrow \ket{3}$ transition is forbidden. The Hamiltonian for this system is given by~\cite{shen2007strongly,roy2017colloquium,shen2005coherent,bradford2012single,witthaut2010photon}:
\begin{equation}
\hat{H} = \hat{H}_p + \hat{H}_a + \hat{H}_i
\label{hamiltonian}
\end{equation}
where 

\begin{align}
\begin{split}
\hat{H}_p &= \intdx \crr^\dagger (x) \left(\omega_0 -i v_g \ddx{}\right) \crr (x)  \\
&+ \intdx \cll^\dagger (x) \left(\omega_0 +i v_g \ddx{}\right)  \cll (x)\\
\hat{H}_a &= \sum_{n=1,2,3}  E_{n}  \ket{n}\bra{n}\\
\hat{H}_i &= \intdx \delta(x)\big( \sum_{n={1,3}} g_n ( \crr^\dagger(x) + \cll^\dagger(x) )\ket{n}\bra{2} + \text{h.c.} \big)
\end{split}
\end{align}

are the photonic, atomic and interaction terms of the Hamiltonian respectively. In this paper we set $\hbar = 1$.  Here, $v_g$ is the group velocity of the photons travelling in the waveguide, $\crr^\dagger (x)$ and $\cll^\dag (x)$ are the creation operators for a right-going and left-going photon at position $x$ in the waveguide. We have chosen to linearize the dispersion relation of the waveguide around the frequency $\omega_0 = E_2 - E_1$. The Hamiltonian of Eq.~\eqref{hamiltonian} describes a three-level atom coupled to a single-mode waveguide. In a three-level atom, the two transitions may couple to different polarizations. For a single-mode waveguide however, as long as the waveguide mode has non-zero components in both polarizations, the waveguide mode will couple to both transitions and our theory will be applicable. Moreover, there are
experimental platforms of three level atoms such as in superconducting qubits~\cite{schuster2005ac} or quantum
dots~\cite{rao2007single}, where these three-level system can be designed to couple with a single-mode waveguide. 

The total excitation number operator $\hat{N}= \int d x c_{L}^{\dagger}(x) c_{L}(x)+\int d x c_{R}^{\dagger}(x) c_{R}(x)+|2\rangle\langle 2|$  commutes with the Hamiltonian $\hat{H}$. The basis states corresponding to the eigenstates of the number operator $\hat{N}$ in the single-excitation regime are:
\begin{equation}
\begin{split}
\{\hat{c}_{R}^{\dagger}(x)|1, v a c\rangle, \hat{c}_{L}^{\dagger}(x)|1, v a c\rangle,\\ \hat{c}_{R}^{\dagger}(x)|3, v a c\rangle, \hat{c}_{L}^{\dagger}(x)|3, v a c\rangle,|2, v a c\rangle \}
\end{split}
\label{basis}
\end{equation}
Since $[H, N] = 0$, we also use these basis states in Eq.~\eqref{basis} to expand the scattering eigenstate of the Hamiltonian with a total energy $E$ in the single-excitation regime as~\cite{shen2005coherent}:
\begin{equation}
\begin{split}
\ket{E} &= \intdx \Big( \left(\phi_{1, R}(x) \hat{c}_{R}^{\dagger}(x)+\phi_{1, L}(x) \hat{c}_{L}^{\dagger}(x)\right)|1, v a c\rangle \\
&+     \left(\phi_{3, R}(x) \hat{c}_{R}^{\dagger}(x)+\phi_{3, L}(x) \hat{c}_{L}^{\dagger}(x)\right)|3, v a c\rangle \Big)  \\
 &+ e_k \ket{2, vac} .
\end{split}
\label{ansatz}
\end{equation}
Here the $\phi$'s are the single-photon field amplitudes. The subscripts of $\phi$ denotes whether the photon is right or left-going, and the state of the atom. For example $\phi_{1, R}$ describes the case where the photon is right-going, and the atom is in state $\ket{1}$.  $e_k$ is the probability amplitude of the three-level atom being in the excited $\ket{2}$ state. $\ket{n,vac}$ is a vacuum state with zero photons and the atom is in state $\ket{n}$.  Using Eq.~\eqref{ansatz} and Eq.~\eqref{hamiltonian} with the time-independent Schrodinger equation $H \ket{E} = E\ket{E}$ gives the following set of linear differential equations:
\begin{align}
E \phi_{1, R}(x) &= - i v_g \frac{\partial \phi_{1, R}(x)}{\partial x}+\phi_{1, R}(x) (E_{1}+\omega_0)+e_{k} g_{1} \delta(x)  \label{diff_eqns_a}
\\
E \phi_{1, L}(x) &=  i v_{g} \frac{\partial \phi_{1, L}(x)}{\partial x}+\phi_{1, L}(x) (E_{1}+\omega_0)+e_{k} g_{1} \delta(x)  \\
E \phi_{3, R}(x) &=  -i v_{g} \frac{\partial \phi_{3, R}(x)}{\partial x}+\phi_{3, R}(x) (E_{3}+\omega_0) +e_{k} g_{3} \delta(x)  \label{diff_eqns_c}\\
E \phi_{3, L}(x)&= i v_{g} \frac{\partial \phi_{3, L}(x)}{\partial x}+\phi_{3, L}(x) (E_{3}+\omega_0) +e_{k} g_{3} \delta(x)  \\
E e_{k} &= E_{2} e_{k}+\sum_{n=1,3} g_{n}\left(\phi_{n, R}(0)+\phi_{n, L}(0)\right).
\label{diff_eqns_b}
\end{align}
We first consider a scenario where the atom is initially in the $\ket{1}$ state and a right-going photon is incoming from the left. Our photon wave-function ansatz is:
\begin{align}
&\phi_{1, R}(x)=e^{i k_{1} x}\left(\Theta(-x)+t_{1} \Theta(x)\right) \\
&\phi_{3, R}(x)=e^{i k_{3} x} t_{3} \Theta(x) \\
&\phi_{1, L}(x)=e^{-i k_{1}^{\prime} x} r_{1} \Theta(-x) \\
&\phi_{3, L}(x)=e^{-i k_{3}^{\prime} x} r_{3} \Theta(-x)
\end{align}
$|t_1|^2, |t_3|^2$ ($|r_1|^2, |r_3|^2$) gives the probability that a photon will be transmitted (reflected) leaving the atom in the $\ket{1}$ or $\ket{3}$ state respectively when the atom was initially in $\ket{1}$. Here, $\Theta(x)$ is the Heaviside function with $\Theta(0) = 0.5$. Using our ansatz and Eq.~\eqref{diff_eqns_a} with $x<0$ and Eq.~\eqref{diff_eqns_c} with $x>0$, we find that $k_1 = (E-E_1-\omega_0)/v_g$, and $k_3 = (E-E_3-\omega_0)/v_g$. Substituting these photon wave-functions into Eq.~\eqref{diff_eqns_a} to~\eqref{diff_eqns_b} gives $k_{1}=k_{1}^{\prime},k_{3}=k_{3}^{\prime}$ as well as the following solutions:
\begin{align}
t_{1}(E) &=1-\frac{\gamma_{1}}{i\left(E_{2}-E\right) +\gamma_{1}+\gamma_{3}}\label{sol_1a} \\
r_{1}(E) &=\frac{-\gamma_{1}}{i\left(E_{2}-E\right) +\gamma_{1}+\gamma_{3}} \\
t_{3}(E) &=\frac{-\sqrt{\gamma_{1} \gamma_{3}}}{i\left(E_{2}-E\right) +\gamma_{1}+\gamma_{3}} \\
r_{3}(E) &=\frac{-\sqrt{\gamma_{1} \gamma_{3}}}{i\left(E_{2}-E\right) +\gamma_{1}+\gamma_{3}}\\
e_{k}(E)&=\frac{-i \sqrt{\gamma_{1} v_{g}}}{i\left(E_{2}-E\right)+\gamma_{1}+\gamma_{3}}\label{sol_1b}
\end{align}
Here $\gamma_{1,3} = g_{1,3}^2/v_g$ are the decay rates of the state $\ket{2}$ into state $\ket{1}$ and state $\ket{3}$ respectively~\cite{rephaeli2012stimulated,witthaut2010photon}. 

In the second scenario, we consider the case where the atom is initially in the $\ket{3}$ state and a right-going photon incoming from the left. Our photon wave-function ansatz is 
\begin{align}
&\phi_{1, R}(x)=e^{i k_{1} x} t_{1}' \Theta(x) \\
&\phi_{3, R}(x)=e^{i k_{3} x}\left(\Theta(-x)+t_{3}' \Theta(x)\right) \\
&\phi_{1, L}(x)=e^{i k_{1}^{\prime} x} r_{1}' \Theta(-x) \\
&\phi_{3, L}(x)=e^{i k_{3}^{\prime} x} r_{3}' \Theta(-x) 
\end{align}
$|t_1'|^2, |t_3'|^2$ ($|r_1'|^2, |r_3'|^2$) gives the probability that a photon will be transmitted (reflected) leaving the atom in the $\ket{1}$ or $\ket{3}$ state respectively when the atom was initially in $\ket{3}$. We can obtain our solutions by substituting this ansatz into the linear differential equations in Eq.~\eqref{diff_eqns_a} to~\eqref{diff_eqns_b} or by simply replacing the labels $1 \leftrightarrow 3$ (due to parity symmetry~\cite{shen2009theory}) in Eq.~\eqref{sol_1a} to Eq.~\eqref{sol_1b}.
\begin{align}
t_{1}'(E) &=\frac{-\sqrt{\gamma_{1} \gamma_{3}}}{i\left(E_{2}-E\right)+\gamma_{1}+\gamma_{3}} \\
r_{1}'(E) &=\frac{-\sqrt{\gamma_{1} \gamma_{3}}}{i\left(E_{2}-E\right) +\gamma_{1}+\gamma_{3}} \\
t_{3}'(E) &=1-\frac{\gamma_{3}}{i\left(E_{2}-E\right) +\gamma_{1}+\gamma_{3}} \\
r_{3}'(E) &=\frac{-\gamma_{3}}{i\left(E_{2}-E\right)+\gamma_{1}+\gamma_{3}} \\
e_{k}^{\prime}(E)&=\frac{-i \sqrt{\gamma_{3} v_{g}}}{i\left(E_{2}-E\right)+\gamma_{1}+\gamma_{3}}
\label{ekp}
\end{align}
We can now summarize the scattering process for a right-going photon of energy $\omega_0 + v_g k$ interacting with a $\Lambda$ atom initially in the $\ket{1}$ state:
\begin{equation}
\begin{split}
\ket{k,1}_R \rightarrow t_1(E) \ket{k,1}_R + t_3 (E) \ket{k-\Delta,3}_R + \\r_1(E) \ket{k,1}_L+ r_3 (E) \ket{k-\Delta,3}_L
\label{twowayscat1}
\end{split}
\end{equation}
where $E= E_1 + v_g k+\omega_0, \Delta = (E_3-E_1)/v_g$, and the label $k$ inside the ket denotes a plane wave of the form $e^{ik x}$. Similarly, for a right-going photon of energy $\omega_0 +v_g (k-\Delta)$ interacting with a $\Lambda$ atom initially in the $\ket{3}$ state:
\begin{equation}
\begin{split}
\ket{k-\Delta,3}_R \rightarrow t_1'(E) \ket{k,1}_R + t_3' (E) \ket{k-\Delta,3}_R + \\
r_1'(E) \ket{k,1}_L + r_3' (E) \ket{k-\Delta,3}_L.
\label{twowayscat2}
\end{split}
\end{equation}
The scattering processes for both cases are shown in Fig.~\ref{maysetup}(b). We note that there are two transmission and reflection output channels for both scenarios.\\

\begin{figure*}[t!]
\centering\includegraphics[width=\textwidth]{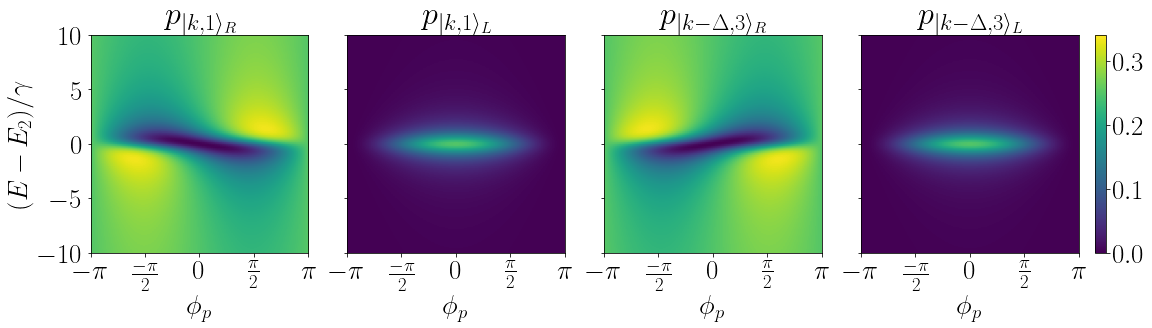}
\caption{Output transmission and reflection for interference terms Eq.~\eqref{int1} to~\eqref{int4} as a function of normalized total energy $(E_2-E)/\gamma$ and initial relative photon phase $\phi_p$ between the two frequency superposition terms. Here, $A= B= C= D = 1/\sqrt{2}, \phi_a= 0, \omega_0/\gamma=10, E_1 = 0, E_2/\gamma = 10, E_3/\gamma = 5,\gamma_1= \gamma_3=\gamma/2$.}
\label{twoway_full}
\end{figure*}
\begin{figure}[h]
\centering
\includegraphics[scale=0.4]{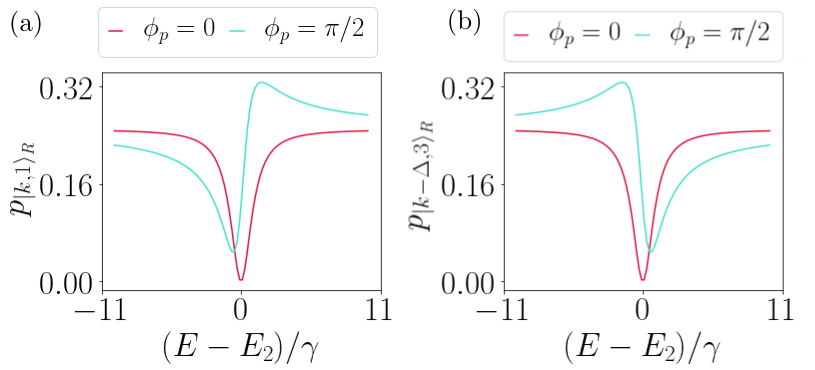}
\caption{Transmission for the scattering output terms (a) $ p_{\ket{k, 1}_R}$ and (b) $ p_{\ket{k-\Delta, 3}_R}$ that exhibit interference as a function of normalized total energy $(E -E_2)/\gamma$ where $A= B= C= D = 1/\sqrt{2}, \phi_a= 0, \omega_0/\gamma=10, E_1 = 0, E_2/\gamma = 10, E_3/\gamma = 5, \gamma_1= \gamma_3=\gamma/2$ for $\phi_p = 0$ and $ \pi/2$.}
\label{trans_fortwo_phis}
\end{figure}

\section{Results}
Let us now consider a separable input state where the photon is in a superposition state $A \ket{k} + Be^{i\phi_p} \ket{k-\Delta}$ which is a frequency superposition state with two frequencies $\omega_0 + v_g k$ and $\omega_0 + v_g (k-\Delta)$ and the atom is in a superposition state of the two ground states $C \ket{1} + D e^{-i\phi_a} \ket{3}$. With our choice of $\omega_0$ the two frequencies are on resonance with the two atomic transitions when $k = 0$.  $A, B, C, D$ are assumed to be real without any loss of generality. $\phi_a$ is the relative phase between the atomic states and $\phi_p$ is the relative phase between the two photon frequencies. Our input state is then made of the tensor product of the photon state $\ket{\psi_p}$ and the atomic state $\ket{\psi_a}$:
\begin{align}
\begin{split}
\left|\psi_{\text {in }}\right\rangle &= \ket{\psi_{p}}\otimes \ket{\psi_{a}}\\
&= (A|k\rangle_R+Be^{i\phi_p}|k-\Delta\rangle_R) \otimes (C|1\rangle+D e^{-i\phi_a}|3\rangle) \\
 &=AC|k,1\rangle_R+ AD e^{-i\phi_a}|k,3\rangle_R+BCe^{i\phi_p}|k-\Delta,1\rangle_R \\
 &+ BDe^{i(\phi_p-\phi_a)}|k-\Delta,3\rangle_R
\label{input}
\end{split}
\end{align}
as depicted in Fig.~\ref{maysetup}(a). The energies of the four components in Eq.~\eqref{input} are $E_1+\omega_0 + v_g k$, $E_1+\omega_0 + v_g(k-\Delta)$, $E_3+\omega_0 + v_gk$ and $E_3+\omega_0 + v_g(k-\Delta)$ respectively. However, with the choice of the two photon frequencies as described above, the $|k,1\rangle_R$ and $|k-\Delta,3\rangle_R$ components in the input state have the same energy $ E_{int} \equiv E_1+\omega_0 + v_g k =E_3+\omega_0+ v_g(k-\Delta)$.


Applying the scattering processes in Eq.~\eqref{twowayscat1} and~\eqref{twowayscat2} to each of the four terms in Eq.~\eqref{input} gives an output state $\left|\phi_{\text {out }}\right\rangle$ with twelve different terms that have the following probabilities:

\begin{align}
p_{\ket{k, 1}_R} & =\left|A C t_1\left(E_{int}\right)+B D e^{i\left(\phi_p-\phi_a\right)} t_1^{\prime}\left(E_{int}\right)\right|^2 \label{int1}\\
p_{\ket{k, 1}_L} & =\left|A C r_1\left(E_{int}\right)+B D e^{i\left(\phi_p-\phi_a\right)} r_1^{\prime}\left(E_{int}\right)\right|^2 \label{int2}\\
p_{\ket{k-\Delta, 3}_R} & =\left|A C t_3\left(E_{int}\right)+B D e^{i\left(\phi_p-\phi_a\right)} t_3^{\prime}\left(E_{int}\right)\right|^2 \label{int3}\\
p_{\ket{k-\Delta, 3}_L} & =\left|A C t_3\left( E_{int}\right)+B D e^{i\left(\phi_p-\phi_a\right)} r_3^{\prime}\left(E_{int}\right)\right|^2 \label{int4}\\
p_{\ket{k+\Delta, 1}_R} & =\left|A D e^{-i \phi_a} t_1^{\prime}\left(E_{+}\right)\right|^2 \label{p5}\\
p_{\ket{k+\Delta, 1}_L} & =\left|A D e^{-i \phi_a} t_1^{\prime}\left(E_{+}\right)\right|^2 \\
p_{\ket{k, 3}_R} & =\left|A D e^{-i \phi_a} t_3^{\prime}\left(E_{+}\right)\right|^2 \\
p_{\ket{k, 3}_L} & =\left|A D e^{-i \phi_a} r_3^{\prime}\left(E_{+}\right)\right|^2 \\
p_{\ket{k-\Delta, 1}_R} & =\left|B C e^{i \phi_p} t_1\left(E_{-}\right)\right|^2 \\
p_{\ket{k-\Delta, 1}_L} & =\left|B C e^{i \phi_p} r_1\left(E_{-}\right)\right|^2 \\
p_{\ket{k-2 \Delta, 3}_R} & =\left|B C e^{i \phi_{p_p}} t_3\left(E_{-}\right)\right|^2 \\
p_{\ket{k-2 \Delta, 3}_L} & =\left|B C e^{i \phi_p} r_3\left(E_{-}\right)\right|^2 .\label{p12}
\end{align}

The subscript of $p$ refers to the outgoing states. For example $p_{\ket{k, 1}^R}$ corresponds to the probability of $|k, 1\rangle_R$. Here, $E_{+}=E_3+\omega_0+v_g k, E_{-}=E_1+\omega_0+v_g(k-\Delta)$. The main result of this paper is that when the input photon frequency states match the atomic resonance levels, both the $|k, 1\rangle_R$ and $|k-\Delta, 3\rangle_R$ input terms can result in the $|k, 1\rangle_R$ output term. Interference between these two channels causes a dependence of the scattering output on the relative phase $(\phi_p - \phi_a)$. A similar logic applies to the $|k-\Delta, 3\rangle_R$ output state. Thus, our scattering output can allow us to compare the effects of both $\phi_a$ and $\phi_p$ in the same platform. This is the advantage of our set-up as this phase information is normally lost when it is squared in intensity measurements. The probability amplitudes in Eq.~\eqref{int1} to~\eqref{int4} can be measured by a joint projective measurements on any of the $\ket{k,1}_R, \ket{k, 1}_L, \ket{k-\Delta, 3}_R$ or $\ket{k-\Delta, 3}_L$ states. Experimentally, the measurement of the photon state can be carried out with a spectrometer~\cite{hempel2013entanglement,reim2011single}. The measurement of the atomic state can then be carried out by scattering a weak classical beam of light against the atom, and by measuring the resulting extinction coefficient~\cite{chen2008memory,hwang2009single}.

Since the total energy is conserved for each term after scattering, we focus only on the subspace with energy $E_{int}$. In Fig.~\ref{twoway_full}, we plot the various probabilities in Eq.~\eqref{int1} to \eqref{int4} as a function of the normalized total energy $(E-E_2)/\gamma$ and the input photon phase $\phi_p$. In Fig.~\ref{trans_fortwo_phis}, we plot a cross-section from Fig.~\ref{twoway_full} at $\phi_p=0,\pi/2$ for the two transmission terms Eq.~\eqref{int1} and Eq.~\eqref{int3} to highlight the $\phi_p$ dependence. Here,
\begin{align}
\begin{split}
A&= B= C= D = 1/\sqrt{2}, \phi_a= 0, \omega_0/\gamma=10, E_1 = 0,\\
 E_2/\gamma &= 10, E_3/\gamma = 5, \gamma_1=\gamma_3=\gamma/2
\label{parameter}
\end{split}
\end{align}
Note that we have picked $\phi_a=0$ but that changing the initial $\phi_a$ only results in a phase shift of the results in Eq.~\eqref{int1} to Eq.~\eqref{int4} (shifting the plots in Fig.~\ref{twoway_full} left or right). Experimentally, a three-level atom can be prepared in a superposition with control over the phase $\phi_a$ in processes such as fractional stimulated Raman adiabatic passage~\cite{vitanov1999creation}. The sum of the panels in Fig.~\ref{twoway_full} do not add to 1 because the non-interference terms must also be included. We see that the reflection terms are symmetric about $\phi_p=0$ while the transmission terms are asymmetric about $\phi_p=0$.  This can be understood by decomposing the input photon state into the scattering mode that maximally excites the atom and the non-scattering mode that does not excite the atom. For arbitrary $\phi_a, \phi_p, A, C$ ($B$ and $D$ are constrained to be $B= \sqrt{1-|A|^2}$ and $D= \sqrt{1-|C|^2}$), and in the sub-space with energy $E_{int}$, we can solve for the non-scattering mode by setting the total excitation probability to be zero:
\begin{equation}
e_{k,\text{total}} = AC (e_k(E_{int}) + BD e^{i(\phi_p-\phi_a)}e_k'(E_{int})) =0
\end{equation}
from which one obtains
\begin{equation}
-AC \sqrt{\gamma_1}=\sqrt{1-|A|^2}\sqrt{1-|C|^2} e^{i(\phi_p-\phi_a)}\sqrt{\gamma_3}.
\end{equation}
For our parameter set in Eq.~\eqref{parameter}, this becomes $-1=e^{i\phi_p}$. Hence, the non-scattering mode $\ket{\psi_n}$ is 
\begin{equation}
\ket{\psi_n}= \frac{1}{\sqrt{2}} \ket{k,1}_R - \frac{1}{\sqrt{2}} \ket{k-\Delta,3}_R .
\end{equation}
Applying the scattering transformations in Eq.~\eqref{twowayscat1} and Eq.~\eqref{twowayscat2} to $\ket{\psi_n}$ gives back the same state:
\begin{equation}
\ket{\psi_n} \rightarrow  \frac{1}{\sqrt{2}} \ket{k,1}_R - \frac{1}{\sqrt{2}} \ket{k-\Delta,3}_R .
\end{equation}
We see that the non-scattering mode has no contribution to the reflection, and its transmission has no frequency dependence. The scattering mode is orthogonal to the non-scattering mode, so for this parameter set in Eq.~\eqref{parameter},  it corresponds to when $\phi_p=0$ and can be written as 
\begin{equation}
\ket{\psi_s}= \frac{1}{\sqrt{2}} \ket{k,1}_R + \frac{1}{\sqrt{2}} \ket{k-\Delta,3}_R .
\end{equation}
The scattering output of $\ket{\psi_s}$ is 
\begin{equation}
\begin{split}
\ket{\psi_s}&\rightarrow  \frac{1}{\sqrt{2}}\Big(\left(1- \frac{\gamma}{i(E_2-E)+\gamma}\right)\ket{k,1}_R \\
&- \frac{\gamma}{i(E_2-E)+\gamma}\ket{k,1}_L \\
 &+\left(1- \frac{\gamma}{i(E_2-E)+\gamma}\right)\ket{k-\Delta,3}_R \\
 &- \frac{\gamma}{i(E_2-E)+\gamma}\ket{k-\Delta,3}_L\Big).
\end{split}
\label{psi_s}
\end{equation}
We see that the transmission coefficient of $\ket{\psi_s}$ is a constant minus a Lorentzian and the reflection is a Lorentzian. 

In the subspace with the energy $E = E_{int}$, any arbitrary input state can be decomposed into a linear superposition of $\ket{\psi_s}$  and $\ket{\psi_n}$, as they also form a basis of the Hilbert space for an incoming right-going photon. In this parameter set of Eq.~\eqref{parameter} when $\phi_p=0$, the contribution is only from $\ket{\psi_s}$. Thus, the transmission is symmetric about $(E_2-E)$, and the reflection is a Lorentzian as expected from Eq.~\eqref{psi_s}. As $\ket{\psi_n}$ does not contribute to the reflection, the reflection remains a Lorentzian for all $\phi_p$. On the other hand when $\phi_p \neq 0,\pm  \pi$, the transmission has components from both $\ket{\psi_s}$ and $\ket{\psi_n}$. The interference of the transmission from the scattering mode $\ket{\psi_s}$ and the transmission from the non-scattering mode $\ket{\psi_n}$ gives rise to a Fano-lineshape~\cite{fan2002analysis}. This explains the asymmetric transmission about $(E_2-E)=0$ for $\phi_p\neq 0, \pm \pi$. The antisymmetry about $\phi_p$ for the transmission plots in Fig.~\ref{twoway_full} is because changing the sign of $\phi_p$ changes the sign of contribution from $\ket{\psi_s}$. 

We note that the existence of a non-scattering input mode that does not excite the atom can be alternatively argued as follows: The atomic excitation $e_{k,\text{total}}$ is a linear function of the amplitudes in the states of $p_{\ket{k, 1}_R}$ and $p_{\ket{k-\Delta, 3}_R}$. Therefore, there is always a particular choice of the amplitudes for which $e_{k,\text{total}} = 0$. 

\section{Conclusion}
In summary, we have shown that the scattering of a single photon in a superposition of two photon frequencies with a $\Lambda$ atom in a superposition of two non-degenerate ground states depends on the initial photon phase between the two photon frequencies (and the initial atomic phase between the ground states). Our results show that three-level $\Lambda$ atoms can be used as a relative photon phase detector for single photons in a superposition state, offering a simpler alternative to other experimental platforms using nonlinear optics~\cite{clemmen2016ramsey} and electro-optic modulators~\cite{lu2020fully}. It also allows us to compare the relative phases of two different objects containing quantum information (the relative phase of the frequency superposition states and the relative phase of the atomic ground states) in the same platform. As applications of quantum information increases, it is of practical relevance to have a platform that can compare the effects of both these phases. Furthermore, as waveguide quantum electrodynamics has a wide range of experimental realizations including superconducting qubits, quantum dots, and cold atoms, the theoretical result in this paper may also allow for more versatility and tunability to make this relative photon phase measurement.

\section*{Acknowledgements}
This work is supported by a Vannevar Bush Faculty Fellowship from the U. S. Department of Defense (Grant No. N00014-17-1-3030). JZ acknowledges A. Poddubny for useful comments.

\section*{Appendix: Effect of dephasing or other losses}
\begin{figure}[h]
\centering
\includegraphics[width=0.45\textwidth]{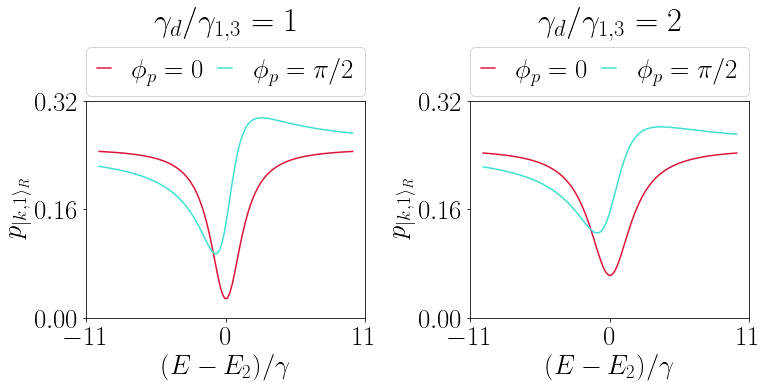}
\caption{Transmission for the scattering output term $ p_{\ket{k, 1}_R}$ as a function of normalized total energy $(E -E_2)/\gamma$ where $A= B= C= D = 1/\sqrt{2}, \phi_a= 0, \omega_0/\gamma=10, E_1 = 0, E_2/\gamma = 10, E_3/\gamma = 5, \gamma_1= \gamma_3=\gamma/2$ for $\phi_p = 0$ and $ \pi/2$. The ratio of the pure dephasing rates over the atomic decay rates are $\gamma_d/\gamma_{1,3}=1,2$ respectively.}
\label{dephasing}
\end{figure}
Various environmental factors such as random fluctuations in the magnetic or electric field, interactions with other particles or quasiparticles or or other defects and coupling with other electromagnetic modes can result in a loss of coherence and hence a decrease in the visibility of transmission or reflectance spectra. In waveguide QED systems, it is typical to phenomenologically add in a pure dephasing rate $\gamma_d$ to account for both dissipative and decoherence effects~\cite{berman2005constraints,liu2020experimental,chan2022quantum,ruostekoski2017arrays} to the atomic decay rates in the transmission and reflection terms.  In Ref.~\cite{shen2009theory}, they use a simple derivation to prove that damping and decoherence effects can be accounted for by making the atomic transition energy complex. For simplicity, we assume the dephasing rate from $\ket{2} \rightarrow \ket{1}$ and $\ket{2} \rightarrow \ket{3}$ are the same. The transmission and reflection coefficients become
\begin{align}
t_{1} &=1-\frac{\gamma_{1}}{i\left(E_{2}-E\right) +\gamma_{1}+\gamma_{3}+\gamma_{d}} \\
r_{1} &=\frac{-\gamma_{1}}{i\left(E_{2}-E\right) +\gamma_{1}+\gamma_{3}+\gamma_{d}} \\
t_{3} &=\frac{-\sqrt{\gamma_{1} \gamma_{3}}}{i\left(E_{2}-E\right) +\gamma_{1}+\gamma_{3}+\gamma_{d}} \\
r_{3} &=\frac{-\sqrt{\gamma_{1} \gamma_{3}}}{i\left(E_{2}-E\right) +\gamma_{1}+\gamma_{3}+\gamma_{d}}\\
e_{k}&=\frac{-i \sqrt{\gamma_{1} v_{g}}}{i\left(E_{2}-E\right)+\gamma_{1}+\gamma_{3}+\gamma_{d}}\\
t_{1}' &=\frac{-\sqrt{\gamma_{1} \gamma_{3}}}{i\left(E_{2}-E\right)+\gamma_{1}+\gamma_{3}+\gamma_{d}} \\
r_{1}' &=\frac{-\sqrt{\gamma_{1} \gamma_{3}}}{i\left(E_{2}-E\right) +\gamma_{1}+\gamma_{3}+\gamma_{d}} \\
t_{3}' &=1-\frac{\gamma_{3}}{i\left(E_{2}-E\right) +\gamma_{1}+\gamma_{3}+\gamma_{d}} \\
r_{3} '&=\frac{-\gamma_{3}}{i\left(E_{2}-E\right)+\gamma_{1}+\gamma_{3}+\gamma_{d}} \\
e_{k}^{\prime}&=\frac{-i \sqrt{\gamma_{3} v_{g}}}{i\left(E_{2}-E\right)+\gamma_{1}+\gamma_{3}+\gamma_{d}}
\end{align}
by adding the dephasing rate in the denominator~\cite{sheremet2021waveguide}. We plot the transmission as in Fig.~\ref{trans_fortwo_phis} for two different dephasing rates in Fig.~\ref{dephasing}. We only include the $p_{|k, 1\rangle_R}$ plot as the dephasing effect on $p_{|k-\Delta, 3\rangle_R}$ is equivalent. We pick the ratio of the pure dephasing rates over the atomic decay rates $\gamma_d/\gamma_{1,3}=1,2$ respectively. We see that when losses are added, the spectra broadens as expected. The peak size of the $\phi_p=0$ transmission dip decreases by $11\%$ and $27\%$ respectively compared to the case with no dephasing rate in Fig.~\ref{trans_fortwo_phis}. In general, the effect that we report in the paper remains the same if $\gamma_d \ll \gamma_{1,3}$.

\nocite{apsrev41Control}
\bibliographystyle{apsrev4}
\bibliography{titleon,bib}

\begin{thebibliography}{46}%
\makeatletter
\providecommand \@ifxundefined [1]{%
 \@ifx{#1\undefined}
}%
\providecommand \@ifnum [1]{%
 \ifnum #1\expandafter \@firstoftwo
 \else \expandafter \@secondoftwo
 \fi
}%
\providecommand \@ifx [1]{%
 \ifx #1\expandafter \@firstoftwo
 \else \expandafter \@secondoftwo
 \fi
}%
\providecommand \natexlab [1]{#1}%
\providecommand \enquote  [1]{``#1''}%
\providecommand \bibnamefont  [1]{#1}%
\providecommand \bibfnamefont [1]{#1}%
\providecommand \citenamefont [1]{#1}%
\providecommand \href@noop [0]{\@secondoftwo}%
\providecommand \href [0]{\begingroup \@sanitize@url \@href}%
\providecommand \@href[1]{\@@startlink{#1}\@@href}%
\providecommand \@@href[1]{\endgroup#1\@@endlink}%
\providecommand \@sanitize@url [0]{\catcode `\\12\catcode `\$12\catcode
  `\&12\catcode `\#12\catcode `\^12\catcode `\_12\catcode `\%12\relax}%
\providecommand \@@startlink[1]{}%
\providecommand \@@endlink[0]{}%
\providecommand \url  [0]{\begingroup\@sanitize@url \@url }%
\providecommand \@url [1]{\endgroup\@href {#1}{\urlprefix }}%
\providecommand \urlprefix  [0]{URL }%
\providecommand \Eprint [0]{\href }%
\providecommand \doibase [0]{http://dx.doi.org/}%
\providecommand \selectlanguage [0]{\@gobble}%
\providecommand \bibinfo  [0]{\@secondoftwo}%
\providecommand \bibfield  [0]{\@secondoftwo}%
\providecommand \translation [1]{[#1]}%
\providecommand \BibitemOpen [0]{}%
\providecommand \bibitemStop [0]{}%
\providecommand \bibitemNoStop [0]{.\EOS\space}%
\providecommand \EOS [0]{\spacefactor3000\relax}%
\providecommand \BibitemShut  [1]{\csname bibitem#1\endcsname}%
\let\auto@bib@innerbib\@empty
\bibitem [{\citenamefont {{Zakka-Bajjani}}\ \emph {et~al.}(2011)\citenamefont
  {{Zakka-Bajjani}}, \citenamefont {{Nguyen}}, \citenamefont {{Lee}},
  \citenamefont {{Vale}}, \citenamefont {{Simmonds}},\ and\ \citenamefont
  {{Aumentado}}}]{zakka2011quantum}%
  \BibitemOpen
  \bibfield  {author} {\bibinfo {author} {\bibfnamefont {E.}~\bibnamefont
  {{Zakka-Bajjani}}}, \bibinfo {author} {\bibfnamefont {F.}~\bibnamefont
  {{Nguyen}}}, \bibinfo {author} {\bibfnamefont {M.}~\bibnamefont {{Lee}}},
  \bibinfo {author} {\bibfnamefont {L.~R.}\ \bibnamefont {{Vale}}}, \bibinfo
  {author} {\bibfnamefont {R.~W.}\ \bibnamefont {{Simmonds}}}, \ and\ \bibinfo
  {author} {\bibfnamefont {J.}~\bibnamefont {{Aumentado}}},\ }\bibfield
  {title} {\enquote {\bibinfo {title} {{Quantum superposition of a single
  microwave photon in two different 'colour' states}},}\ }\href {\doibase
  10.1038/nphys2035} {\bibfield  {journal} {\bibinfo  {journal} {Nature
  Physics}\ }\textbf {\bibinfo {volume} {7}},\ \bibinfo {pages} {599--603}
  (\bibinfo {year} {2011})}\BibitemShut {NoStop}%
\bibitem [{\citenamefont {Clemmen}\ \emph {et~al.}(2016)\citenamefont
  {Clemmen}, \citenamefont {Farsi}, \citenamefont {Ramelow},\ and\
  \citenamefont {Gaeta}}]{clemmen2016ramsey}%
  \BibitemOpen
  \bibfield  {author} {\bibinfo {author} {\bibfnamefont {S.}~\bibnamefont
  {Clemmen}}, \bibinfo {author} {\bibfnamefont {A.}~\bibnamefont {Farsi}},
  \bibinfo {author} {\bibfnamefont {S.}~\bibnamefont {Ramelow}}, \ and\
  \bibinfo {author} {\bibfnamefont {A.~L.}\ \bibnamefont {Gaeta}},\ }\bibfield
  {title} {\enquote {\bibinfo {title} {Ramsey interference with single
  photons},}\ }\href {\doibase 10.1103/PhysRevLett.117.223601} {\bibfield
  {journal} {\bibinfo  {journal} {Phys. Rev. Lett.}\ }\textbf {\bibinfo
  {volume} {117}},\ \bibinfo {pages} {223601} (\bibinfo {year}
  {2016})}\BibitemShut {NoStop}%
\bibitem [{\citenamefont {{Aguayo-Alvarado}}\ \emph {et~al.}(2022)\citenamefont
  {{Aguayo-Alvarado}}, \citenamefont {{Dom{\'\i}nguez-Serna}}, \citenamefont
  {{De La Cruz}},\ and\ \citenamefont
  {{Garay-Palmett}}}]{aguayo2022integrated}%
  \BibitemOpen
  \bibfield  {author} {\bibinfo {author} {\bibfnamefont {A.~L.}\ \bibnamefont
  {{Aguayo-Alvarado}}}, \bibinfo {author} {\bibfnamefont {F.}~\bibnamefont
  {{Dom{\'\i}nguez-Serna}}}, \bibinfo {author} {\bibfnamefont {W.}~\bibnamefont
  {{De La Cruz}}}, \ and\ \bibinfo {author} {\bibfnamefont {K.}~\bibnamefont
  {{Garay-Palmett}}},\ }\bibfield  {title} {\enquote {\bibinfo {title} {{An
  integrated photonic circuit for color qubit preparation by third-order
  nonlinear interactions}},}\ }\href {\doibase 10.1038/s41598-022-09116-w}
  {\bibfield  {journal} {\bibinfo  {journal} {Scientific Reports}\ }\textbf
  {\bibinfo {volume} {12}},\ \bibinfo {eid} {5154} (\bibinfo {year}
  {2022})}\BibitemShut {NoStop}%
\bibitem [{\citenamefont {{Tanzilli}}\ \emph {et~al.}(2005)\citenamefont
  {{Tanzilli}}, \citenamefont {{Tittel}}, \citenamefont {{Halder}},
  \citenamefont {{Alibart}}, \citenamefont {{Baldi}}, \citenamefont {{Gisin}},\
  and\ \citenamefont {{Zbinden}}}]{tanzilli2005photonic}%
  \BibitemOpen
  \bibfield  {author} {\bibinfo {author} {\bibfnamefont {S.}~\bibnamefont
  {{Tanzilli}}}, \bibinfo {author} {\bibfnamefont {W.}~\bibnamefont
  {{Tittel}}}, \bibinfo {author} {\bibfnamefont {M.}~\bibnamefont {{Halder}}},
  \bibinfo {author} {\bibfnamefont {O.}~\bibnamefont {{Alibart}}}, \bibinfo
  {author} {\bibfnamefont {P.}~\bibnamefont {{Baldi}}}, \bibinfo {author}
  {\bibfnamefont {N.}~\bibnamefont {{Gisin}}}, \ and\ \bibinfo {author}
  {\bibfnamefont {H.}~\bibnamefont {{Zbinden}}},\ }\bibfield  {title} {\enquote
  {\bibinfo {title} {{A photonic quantum information interface}},}\ }\href
  {\doibase 10.1038/nature04009} {\bibfield  {journal} {\bibinfo  {journal}
  {\nat}\ }\textbf {\bibinfo {volume} {437}},\ \bibinfo {pages} {116--120}
  (\bibinfo {year} {2005})}\BibitemShut {NoStop}%
\bibitem [{\citenamefont {Karnieli}\ and\ \citenamefont
  {Arie}(2018)}]{karnieli2018frequency}%
  \BibitemOpen
  \bibfield  {author} {\bibinfo {author} {\bibfnamefont {A.}~\bibnamefont
  {Karnieli}}\ and\ \bibinfo {author} {\bibfnamefont {A.}~\bibnamefont
  {Arie}},\ }\bibfield  {title} {\enquote {\bibinfo {title} {Frequency domain
  stern-gerlach effect for photonic qubits and qutrits},}\ }\href
  {http://opg.optica.org/optica/abstract.cfm?URI=optica-5-10-1297} {\bibfield
  {journal} {\bibinfo  {journal} {Optica}\ }\textbf {\bibinfo {volume} {5}},\
  \bibinfo {pages} {1297--1303} (\bibinfo {year} {2018})}\BibitemShut {NoStop}%
\bibitem [{\citenamefont {{Lukens}}\ and\ \citenamefont
  {{Lougovski}}(2017)}]{lukens2017frequency}%
  \BibitemOpen
  \bibfield  {author} {\bibinfo {author} {\bibfnamefont {J.~M.}\ \bibnamefont
  {{Lukens}}}\ and\ \bibinfo {author} {\bibfnamefont {P.}~\bibnamefont
  {{Lougovski}}},\ }\bibfield  {title} {\enquote {\bibinfo {title}
  {{Frequency-encoded photonic qubits for scalable quantum information
  processing}},}\ }\href {\doibase 10.1364/OPTICA.4.000008} {\bibfield
  {journal} {\bibinfo  {journal} {Optica}\ }\textbf {\bibinfo {volume} {4}},\
  \bibinfo {pages} {8} (\bibinfo {year} {2017})}\BibitemShut {NoStop}%
\bibitem [{\citenamefont {Lu}\ \emph {et~al.}(2020)\citenamefont {Lu},
  \citenamefont {Simmerman}, \citenamefont {Lougovski}, \citenamefont
  {Weiner},\ and\ \citenamefont {Lukens}}]{lu2020fully}%
  \BibitemOpen
  \bibfield  {author} {\bibinfo {author} {\bibfnamefont {H.-H.}\ \bibnamefont
  {Lu}}, \bibinfo {author} {\bibfnamefont {E.~M.}\ \bibnamefont {Simmerman}},
  \bibinfo {author} {\bibfnamefont {P.}~\bibnamefont {Lougovski}}, \bibinfo
  {author} {\bibfnamefont {A.~M.}\ \bibnamefont {Weiner}}, \ and\ \bibinfo
  {author} {\bibfnamefont {J.~M.}\ \bibnamefont {Lukens}},\ }\bibfield  {title}
  {\enquote {\bibinfo {title} {Fully arbitrary control of frequency-bin
  qubits},}\ }\href {\doibase 10.1103/PhysRevLett.125.120503} {\bibfield
  {journal} {\bibinfo  {journal} {Phys. Rev. Lett.}\ }\textbf {\bibinfo
  {volume} {125}},\ \bibinfo {pages} {120503} (\bibinfo {year}
  {2020})}\BibitemShut {NoStop}%
\bibitem [{\citenamefont {Lu}\ \emph {et~al.}(2019)\citenamefont {Lu},
  \citenamefont {Lukens}, \citenamefont {Williams}, \citenamefont {Imany},
  \citenamefont {Peters}, \citenamefont {Weiner},\ and\ \citenamefont
  {Lougovski}}]{lu2019controlled}%
  \BibitemOpen
  \bibfield  {author} {\bibinfo {author} {\bibfnamefont {H.-H.}\ \bibnamefont
  {Lu}}, \bibinfo {author} {\bibfnamefont {J.~M.}\ \bibnamefont {Lukens}},
  \bibinfo {author} {\bibfnamefont {B.~P.}\ \bibnamefont {Williams}}, \bibinfo
  {author} {\bibfnamefont {P.}~\bibnamefont {Imany}}, \bibinfo {author}
  {\bibfnamefont {N.~A.}\ \bibnamefont {Peters}}, \bibinfo {author}
  {\bibfnamefont {A.~M.}\ \bibnamefont {Weiner}}, \ and\ \bibinfo {author}
  {\bibfnamefont {P.}~\bibnamefont {Lougovski}},\ }\bibfield  {title} {\enquote
  {\bibinfo {title} {A controlled-not gate for frequency-bin qubits},}\ }\href
  {https://www.nature.com/articles/s41534-019-0137-z} {\bibfield  {journal}
  {\bibinfo  {journal} {npj Quantum Information}\ }\textbf {\bibinfo {volume}
  {5}},\ \bibinfo {pages} {1--8} (\bibinfo {year} {2019})}\BibitemShut
  {NoStop}%
\bibitem [{\citenamefont {Lu}\ \emph {et~al.}(2018)\citenamefont {Lu},
  \citenamefont {Lukens}, \citenamefont {Peters}, \citenamefont {Williams},
  \citenamefont {Weiner},\ and\ \citenamefont {Lougovski}}]{lu2018quantum}%
  \BibitemOpen
  \bibfield  {author} {\bibinfo {author} {\bibfnamefont {H.-H.}\ \bibnamefont
  {Lu}}, \bibinfo {author} {\bibfnamefont {J.~M.}\ \bibnamefont {Lukens}},
  \bibinfo {author} {\bibfnamefont {N.~A.}\ \bibnamefont {Peters}}, \bibinfo
  {author} {\bibfnamefont {B.~P.}\ \bibnamefont {Williams}}, \bibinfo {author}
  {\bibfnamefont {A.~M.}\ \bibnamefont {Weiner}}, \ and\ \bibinfo {author}
  {\bibfnamefont {P.}~\bibnamefont {Lougovski}},\ }\bibfield  {title} {\enquote
  {\bibinfo {title} {Quantum interference and correlation control of
  frequency-bin qubits},}\ }\href {\doibase 10.1364/OPTICA.5.001455} {\bibfield
   {journal} {\bibinfo  {journal} {Optica}\ }\textbf {\bibinfo {volume} {5}},\
  \bibinfo {pages} {1455--1460} (\bibinfo {year} {2018})}\BibitemShut {NoStop}%
\bibitem [{\citenamefont {Olislager}\ \emph {et~al.}(2010)\citenamefont
  {Olislager}, \citenamefont {Cussey}, \citenamefont {Nguyen}, \citenamefont
  {Emplit}, \citenamefont {Massar}, \citenamefont {Merolla},\ and\
  \citenamefont {Huy}}]{olisager2010frequency}%
  \BibitemOpen
  \bibfield  {author} {\bibinfo {author} {\bibfnamefont {L.}~\bibnamefont
  {Olislager}}, \bibinfo {author} {\bibfnamefont {J.}~\bibnamefont {Cussey}},
  \bibinfo {author} {\bibfnamefont {A.~T.}\ \bibnamefont {Nguyen}}, \bibinfo
  {author} {\bibfnamefont {P.}~\bibnamefont {Emplit}}, \bibinfo {author}
  {\bibfnamefont {S.}~\bibnamefont {Massar}}, \bibinfo {author} {\bibfnamefont
  {J.-M.}\ \bibnamefont {Merolla}}, \ and\ \bibinfo {author} {\bibfnamefont
  {K.~P.}\ \bibnamefont {Huy}},\ }\bibfield  {title} {\enquote {\bibinfo
  {title} {Frequency-bin entangled photons},}\ }\href {\doibase
  10.1103/PhysRevA.82.013804} {\bibfield  {journal} {\bibinfo  {journal} {Phys.
  Rev. A}\ }\textbf {\bibinfo {volume} {82}},\ \bibinfo {pages} {013804}
  (\bibinfo {year} {2010})}\BibitemShut {NoStop}%
\bibitem [{\citenamefont {Huang}\ and\ \citenamefont
  {Kumar}(1992)}]{huang1992observation}%
  \BibitemOpen
  \bibfield  {author} {\bibinfo {author} {\bibfnamefont {J.}~\bibnamefont
  {Huang}}\ and\ \bibinfo {author} {\bibfnamefont {P.}~\bibnamefont {Kumar}},\
  }\bibfield  {title} {\enquote {\bibinfo {title} {Observation of quantum
  frequency conversion},}\ }\href {\doibase 10.1103/PhysRevLett.68.2153}
  {\bibfield  {journal} {\bibinfo  {journal} {Phys. Rev. Lett.}\ }\textbf
  {\bibinfo {volume} {68}},\ \bibinfo {pages} {2153--2156} (\bibinfo {year}
  {1992})}\BibitemShut {NoStop}%
\bibitem [{\citenamefont {{Kobayashi}}\ \emph {et~al.}(2016)\citenamefont
  {{Kobayashi}}, \citenamefont {{Ikuta}}, \citenamefont {{Yasui}},
  \citenamefont {{Miki}}, \citenamefont {{Yamashita}}, \citenamefont {{Terai}},
  \citenamefont {{Yamamoto}}, \citenamefont {{Koashi}},\ and\ \citenamefont
  {{Imoto}}}]{kobayashi2016frequency}%
  \BibitemOpen
  \bibfield  {author} {\bibinfo {author} {\bibfnamefont {T.}~\bibnamefont
  {{Kobayashi}}}, \bibinfo {author} {\bibfnamefont {R.}~\bibnamefont
  {{Ikuta}}}, \bibinfo {author} {\bibfnamefont {S.}~\bibnamefont {{Yasui}}},
  \bibinfo {author} {\bibfnamefont {S.}~\bibnamefont {{Miki}}}, \bibinfo
  {author} {\bibfnamefont {T.}~\bibnamefont {{Yamashita}}}, \bibinfo {author}
  {\bibfnamefont {H.}~\bibnamefont {{Terai}}}, \bibinfo {author} {\bibfnamefont
  {T.}~\bibnamefont {{Yamamoto}}}, \bibinfo {author} {\bibfnamefont
  {M.}~\bibnamefont {{Koashi}}}, \ and\ \bibinfo {author} {\bibfnamefont
  {N.}~\bibnamefont {{Imoto}}},\ }\bibfield  {title} {\enquote {\bibinfo
  {title} {{Frequency-domain Hong-Ou-Mandel interference}},}\ }\href {\doibase
  10.1038/nphoton.2016.74} {\bibfield  {journal} {\bibinfo  {journal} {Nature
  Photonics}\ }\textbf {\bibinfo {volume} {10}},\ \bibinfo {pages} {441--444}
  (\bibinfo {year} {2016})}\BibitemShut {NoStop}%
\bibitem [{\citenamefont {Kobayashi}\ \emph {et~al.}(2017)\citenamefont
  {Kobayashi}, \citenamefont {Yamazaki}, \citenamefont {Matsuki}, \citenamefont
  {Ikuta}, \citenamefont {Miki}, \citenamefont {Yamashita}, \citenamefont
  {Terai}, \citenamefont {Yamamoto}, \citenamefont {Koashi},\ and\
  \citenamefont {Imoto}}]{kobayashi2017coherent}%
  \BibitemOpen
  \bibfield  {author} {\bibinfo {author} {\bibfnamefont {T.}~\bibnamefont
  {Kobayashi}}, \bibinfo {author} {\bibfnamefont {D.}~\bibnamefont {Yamazaki}},
  \bibinfo {author} {\bibfnamefont {K.}~\bibnamefont {Matsuki}}, \bibinfo
  {author} {\bibfnamefont {R.}~\bibnamefont {Ikuta}}, \bibinfo {author}
  {\bibfnamefont {S.}~\bibnamefont {Miki}}, \bibinfo {author} {\bibfnamefont
  {T.}~\bibnamefont {Yamashita}}, \bibinfo {author} {\bibfnamefont
  {H.}~\bibnamefont {Terai}}, \bibinfo {author} {\bibfnamefont
  {T.}~\bibnamefont {Yamamoto}}, \bibinfo {author} {\bibfnamefont
  {M.}~\bibnamefont {Koashi}}, \ and\ \bibinfo {author} {\bibfnamefont
  {N.}~\bibnamefont {Imoto}},\ }\bibfield  {title} {\enquote {\bibinfo {title}
  {Mach-zehnder interferometer using frequency-domain beamsplitter},}\ }\href
  {\doibase 10.1364/OE.25.012052} {\bibfield  {journal} {\bibinfo  {journal}
  {Opt. Express}\ }\textbf {\bibinfo {volume} {25}},\ \bibinfo {pages}
  {12052--12060} (\bibinfo {year} {2017})}\BibitemShut {NoStop}%
\bibitem [{\citenamefont {McGuinness}\ \emph {et~al.}(2010)\citenamefont
  {McGuinness}, \citenamefont {Raymer}, \citenamefont {McKinstrie},\ and\
  \citenamefont {Radic}}]{mcguiness2010quantum}%
  \BibitemOpen
  \bibfield  {author} {\bibinfo {author} {\bibfnamefont {H.~J.}\ \bibnamefont
  {McGuinness}}, \bibinfo {author} {\bibfnamefont {M.~G.}\ \bibnamefont
  {Raymer}}, \bibinfo {author} {\bibfnamefont {C.~J.}\ \bibnamefont
  {McKinstrie}}, \ and\ \bibinfo {author} {\bibfnamefont {S.}~\bibnamefont
  {Radic}},\ }\bibfield  {title} {\enquote {\bibinfo {title} {Quantum frequency
  translation of single-photon states in a photonic crystal fiber},}\ }\href
  {\doibase 10.1103/PhysRevLett.105.093604} {\bibfield  {journal} {\bibinfo
  {journal} {Phys. Rev. Lett.}\ }\textbf {\bibinfo {volume} {105}},\ \bibinfo
  {pages} {093604} (\bibinfo {year} {2010})}\BibitemShut {NoStop}%
\bibitem [{\citenamefont {Dmitriev}\ \emph {et~al.}(2017)\citenamefont
  {Dmitriev}, \citenamefont {Shaikhaidarov}, \citenamefont {Antonov},
  \citenamefont {H{\"o}nigl-Decrinis},\ and\ \citenamefont
  {Astafiev}}]{dmitriev2017quantum}%
  \BibitemOpen
  \bibfield  {author} {\bibinfo {author} {\bibfnamefont {A.~Y.}\ \bibnamefont
  {Dmitriev}}, \bibinfo {author} {\bibfnamefont {R.}~\bibnamefont
  {Shaikhaidarov}}, \bibinfo {author} {\bibfnamefont {V.}~\bibnamefont
  {Antonov}}, \bibinfo {author} {\bibfnamefont {T.}~\bibnamefont
  {H{\"o}nigl-Decrinis}}, \ and\ \bibinfo {author} {\bibfnamefont
  {O.}~\bibnamefont {Astafiev}},\ }\bibfield  {title} {\enquote {\bibinfo
  {title} {Quantum wave mixing and visualisation of coherent and superposed
  photonic states in a waveguide},}\ }\href
  {https://www.nature.com/articles/s41467-017-01471-x} {\bibfield  {journal}
  {\bibinfo  {journal} {Nature communications}\ }\textbf {\bibinfo {volume}
  {8}},\ \bibinfo {pages} {1--6} (\bibinfo {year} {2017})}\BibitemShut
  {NoStop}%
\bibitem [{\citenamefont {{Knill}}\ \emph {et~al.}(2001)\citenamefont
  {{Knill}}, \citenamefont {{Laflamme}},\ and\ \citenamefont
  {{Milburn}}}]{knill2001scheme}%
  \BibitemOpen
  \bibfield  {author} {\bibinfo {author} {\bibfnamefont {E.}~\bibnamefont
  {{Knill}}}, \bibinfo {author} {\bibfnamefont {R.}~\bibnamefont {{Laflamme}}},
  \ and\ \bibinfo {author} {\bibfnamefont {G.~J.}\ \bibnamefont {{Milburn}}},\
  }\bibfield  {title} {\enquote {\bibinfo {title} {{A scheme for efficient
  quantum computation with linear optics}},}\ }\href {\doibase
  10.1038/35051009} {\bibfield  {journal} {\bibinfo  {journal} {\nat}\ }\textbf
  {\bibinfo {volume} {409}},\ \bibinfo {pages} {46--52} (\bibinfo {year}
  {2001})}\BibitemShut {NoStop}%
\bibitem [{\citenamefont {Javid}\ \emph {et~al.}(2021)\citenamefont {Javid},
  \citenamefont {Ling}, \citenamefont {Staffa}, \citenamefont {Li},
  \citenamefont {He},\ and\ \citenamefont {Lin}}]{ultrabroadband2021javid}%
  \BibitemOpen
  \bibfield  {author} {\bibinfo {author} {\bibfnamefont {U.~A.}\ \bibnamefont
  {Javid}}, \bibinfo {author} {\bibfnamefont {J.}~\bibnamefont {Ling}},
  \bibinfo {author} {\bibfnamefont {J.}~\bibnamefont {Staffa}}, \bibinfo
  {author} {\bibfnamefont {M.}~\bibnamefont {Li}}, \bibinfo {author}
  {\bibfnamefont {Y.}~\bibnamefont {He}}, \ and\ \bibinfo {author}
  {\bibfnamefont {Q.}~\bibnamefont {Lin}},\ }\bibfield  {title} {\enquote
  {\bibinfo {title} {Ultrabroadband entangled photons on a nanophotonic
  chip},}\ }\href {\doibase 10.1103/PhysRevLett.127.183601} {\bibfield
  {journal} {\bibinfo  {journal} {Phys. Rev. Lett.}\ }\textbf {\bibinfo
  {volume} {127}},\ \bibinfo {pages} {183601} (\bibinfo {year}
  {2021})}\BibitemShut {NoStop}%
\bibitem [{\citenamefont {{Giovannetti}}\ \emph {et~al.}(2011)\citenamefont
  {{Giovannetti}}, \citenamefont {{Lloyd}},\ and\ \citenamefont
  {{Maccone}}}]{giovannetti2011advances}%
  \BibitemOpen
  \bibfield  {author} {\bibinfo {author} {\bibfnamefont {V.}~\bibnamefont
  {{Giovannetti}}}, \bibinfo {author} {\bibfnamefont {S.}~\bibnamefont
  {{Lloyd}}}, \ and\ \bibinfo {author} {\bibfnamefont {L.}~\bibnamefont
  {{Maccone}}},\ }\bibfield  {title} {\enquote {\bibinfo {title} {{Advances in
  quantum metrology}},}\ }\href {\doibase 10.1038/nphoton.2011.35} {\bibfield
  {journal} {\bibinfo  {journal} {Nature Photonics}\ }\textbf {\bibinfo
  {volume} {5}},\ \bibinfo {pages} {222--229} (\bibinfo {year} {2011})},\
  \Eprint {http://arxiv.org/abs/1102.2318} {arXiv:1102.2318 [quant-ph]}
  \BibitemShut {NoStop}%
\bibitem [{\citenamefont {Rinehart}\ \emph {et~al.}(2012)\citenamefont
  {Rinehart}, \citenamefont {Zhu},\ and\ \citenamefont
  {Wax}}]{rinehart2012quantitative}%
  \BibitemOpen
  \bibfield  {author} {\bibinfo {author} {\bibfnamefont {M.}~\bibnamefont
  {Rinehart}}, \bibinfo {author} {\bibfnamefont {Y.}~\bibnamefont {Zhu}}, \
  and\ \bibinfo {author} {\bibfnamefont {A.}~\bibnamefont {Wax}},\ }\bibfield
  {title} {\enquote {\bibinfo {title} {Quantitative phase spectroscopy},}\
  }\href {https://opg.optica.org/boe/fulltext.cfm?uri=boe-3-5-958&id=231914}
  {\bibfield  {journal} {\bibinfo  {journal} {Biomedical optics express}\
  }\textbf {\bibinfo {volume} {3}},\ \bibinfo {pages} {958--965} (\bibinfo
  {year} {2012})}\BibitemShut {NoStop}%
\bibitem [{\citenamefont {Yesharim}\ \emph {et~al.}(2022)\citenamefont
  {Yesharim}, \citenamefont {Karnieli}, \citenamefont {Jackel}, \citenamefont
  {Di~Domenico}, \citenamefont {Trajtenberg-Mills},\ and\ \citenamefont
  {Arie}}]{yesharim2022observation}%
  \BibitemOpen
  \bibfield  {author} {\bibinfo {author} {\bibfnamefont {O.}~\bibnamefont
  {Yesharim}}, \bibinfo {author} {\bibfnamefont {A.}~\bibnamefont {Karnieli}},
  \bibinfo {author} {\bibfnamefont {S.}~\bibnamefont {Jackel}}, \bibinfo
  {author} {\bibfnamefont {G.}~\bibnamefont {Di~Domenico}}, \bibinfo {author}
  {\bibfnamefont {S.}~\bibnamefont {Trajtenberg-Mills}}, \ and\ \bibinfo
  {author} {\bibfnamefont {A.}~\bibnamefont {Arie}},\ }\bibfield  {title}
  {\enquote {\bibinfo {title} {Observation of the all-optical stern--gerlach
  effect in nonlinear optics},}\ }\href@noop {} {\bibfield  {journal} {\bibinfo
   {journal} {Nature Photonics}\ ,\ \bibinfo {pages} {1--6}} (\bibinfo {year}
  {2022})}\BibitemShut {NoStop}%
\bibitem [{\citenamefont {Sheremet}\ \emph {et~al.}(2023)\citenamefont
  {Sheremet}, \citenamefont {Petrov}, \citenamefont {Iorsh}, \citenamefont
  {Poshakinskiy},\ and\ \citenamefont {Poddubny}}]{sheremet2021waveguide}%
  \BibitemOpen
  \bibfield  {author} {\bibinfo {author} {\bibfnamefont {A.~S.}\ \bibnamefont
  {Sheremet}}, \bibinfo {author} {\bibfnamefont {M.~I.}\ \bibnamefont
  {Petrov}}, \bibinfo {author} {\bibfnamefont {I.~V.}\ \bibnamefont {Iorsh}},
  \bibinfo {author} {\bibfnamefont {A.~V.}\ \bibnamefont {Poshakinskiy}}, \
  and\ \bibinfo {author} {\bibfnamefont {A.~N.}\ \bibnamefont {Poddubny}},\
  }\bibfield  {title} {\enquote {\bibinfo {title} {Waveguide quantum
  electrodynamics: Collective radiance and photon-photon correlations},}\
  }\href {\doibase 10.1103/RevModPhys.95.015002} {\bibfield  {journal}
  {\bibinfo  {journal} {Rev. Mod. Phys.}\ }\textbf {\bibinfo {volume} {95}},\
  \bibinfo {pages} {015002} (\bibinfo {year} {2023})}\BibitemShut {NoStop}%
\bibitem [{\citenamefont {Roy}\ \emph {et~al.}(2017)\citenamefont {Roy},
  \citenamefont {Wilson},\ and\ \citenamefont
  {Firstenberg}}]{roy2017colloquium}%
  \BibitemOpen
  \bibfield  {author} {\bibinfo {author} {\bibfnamefont {D.}~\bibnamefont
  {Roy}}, \bibinfo {author} {\bibfnamefont {C.~M.}\ \bibnamefont {Wilson}}, \
  and\ \bibinfo {author} {\bibfnamefont {O.}~\bibnamefont {Firstenberg}},\
  }\bibfield  {title} {\enquote {\bibinfo {title} {Colloquium: Strongly
  interacting photons in one-dimensional continuum},}\ }\href {\doibase
  10.1103/RevModPhys.89.021001} {\bibfield  {journal} {\bibinfo  {journal}
  {Rev. Mod. Phys.}\ }\textbf {\bibinfo {volume} {89}},\ \bibinfo {pages}
  {021001} (\bibinfo {year} {2017})}\BibitemShut {NoStop}%
\bibitem [{\citenamefont {Türschmann}\ \emph {et~al.}(2019)\citenamefont
  {Türschmann}, \citenamefont {Jeannic}, \citenamefont {Simonsen},
  \citenamefont {Haakh}, \citenamefont {Götzinger}, \citenamefont
  {Sandoghdar}, \citenamefont {Lodahl},\ and\ \citenamefont
  {Rotenberg}}]{turschmann2019coherent}%
  \BibitemOpen
  \bibfield  {author} {\bibinfo {author} {\bibfnamefont {P.}~\bibnamefont
  {Türschmann}}, \bibinfo {author} {\bibfnamefont {H.~L.}\ \bibnamefont
  {Jeannic}}, \bibinfo {author} {\bibfnamefont {S.~F.}\ \bibnamefont
  {Simonsen}}, \bibinfo {author} {\bibfnamefont {H.~R.}\ \bibnamefont {Haakh}},
  \bibinfo {author} {\bibfnamefont {S.}~\bibnamefont {Götzinger}}, \bibinfo
  {author} {\bibfnamefont {V.}~\bibnamefont {Sandoghdar}}, \bibinfo {author}
  {\bibfnamefont {P.}~\bibnamefont {Lodahl}}, \ and\ \bibinfo {author}
  {\bibfnamefont {N.}~\bibnamefont {Rotenberg}},\ }\bibfield  {title} {\enquote
  {\bibinfo {title} {Coherent nonlinear optics of quantum emitters in
  nanophotonic waveguides},}\ }\href {\doibase doi:10.1515/nanoph-2019-0126}
  {\bibfield  {journal} {\bibinfo  {journal} {Nanophotonics}\ }\textbf
  {\bibinfo {volume} {8}},\ \bibinfo {pages} {1641--1657} (\bibinfo {year}
  {2019})}\BibitemShut {NoStop}%
\bibitem [{\citenamefont {Chang}\ \emph {et~al.}(2018)\citenamefont {Chang},
  \citenamefont {Douglas}, \citenamefont {Gonz\'alez-Tudela}, \citenamefont
  {Hung},\ and\ \citenamefont {Kimble}}]{chang2018colloquium}%
  \BibitemOpen
  \bibfield  {author} {\bibinfo {author} {\bibfnamefont {D.~E.}\ \bibnamefont
  {Chang}}, \bibinfo {author} {\bibfnamefont {J.~S.}\ \bibnamefont {Douglas}},
  \bibinfo {author} {\bibfnamefont {A.}~\bibnamefont {Gonz\'alez-Tudela}},
  \bibinfo {author} {\bibfnamefont {C.-L.}\ \bibnamefont {Hung}}, \ and\
  \bibinfo {author} {\bibfnamefont {H.~J.}\ \bibnamefont {Kimble}},\ }\bibfield
   {title} {\enquote {\bibinfo {title} {Colloquium: Quantum matter built from
  nanoscopic lattices of atoms and photons},}\ }\href {\doibase
  10.1103/RevModPhys.90.031002} {\bibfield  {journal} {\bibinfo  {journal}
  {Rev. Mod. Phys.}\ }\textbf {\bibinfo {volume} {90}},\ \bibinfo {pages}
  {031002} (\bibinfo {year} {2018})}\BibitemShut {NoStop}%
\bibitem [{\citenamefont {Djotyan}\ \emph {et~al.}(2008)\citenamefont
  {Djotyan}, \citenamefont {Bakos}, \citenamefont {Demeter}, \citenamefont
  {S\"{o}rlei}, \citenamefont {Szigeti},\ and\ \citenamefont
  {Dzsotjan}}]{djotyan2008multiphoton}%
  \BibitemOpen
  \bibfield  {author} {\bibinfo {author} {\bibfnamefont {G.~P.}\ \bibnamefont
  {Djotyan}}, \bibinfo {author} {\bibfnamefont {J.~S.}\ \bibnamefont {Bakos}},
  \bibinfo {author} {\bibfnamefont {G.}~\bibnamefont {Demeter}}, \bibinfo
  {author} {\bibfnamefont {Z.}~\bibnamefont {S\"{o}rlei}}, \bibinfo {author}
  {\bibfnamefont {J.}~\bibnamefont {Szigeti}}, \ and\ \bibinfo {author}
  {\bibfnamefont {D.}~\bibnamefont {Dzsotjan}},\ }\bibfield  {title} {\enquote
  {\bibinfo {title} {Creation of a coherent superposition of quantum states by
  a single frequency-chirped short laser pulse},}\ }\href {\doibase
  10.1364/JOSAB.25.000166} {\bibfield  {journal} {\bibinfo  {journal} {J. Opt.
  Soc. Am. B}\ }\textbf {\bibinfo {volume} {25}},\ \bibinfo {pages} {166--174}
  (\bibinfo {year} {2008})}\BibitemShut {NoStop}%
\bibitem [{\citenamefont {{He}}\ \emph {et~al.}(2009)\citenamefont {{He}},
  \citenamefont {{Zhang}}, \citenamefont {{Zhou}}, \citenamefont {{Chen}},\
  and\ \citenamefont {{Guo}}}]{he2009arbitrary}%
  \BibitemOpen
  \bibfield  {author} {\bibinfo {author} {\bibfnamefont {J.}~\bibnamefont
  {{He}}}, \bibinfo {author} {\bibfnamefont {Y.-S.}\ \bibnamefont {{Zhang}}},
  \bibinfo {author} {\bibfnamefont {X.-F.}\ \bibnamefont {{Zhou}}}, \bibinfo
  {author} {\bibfnamefont {Q.-F.}\ \bibnamefont {{Chen}}}, \ and\ \bibinfo
  {author} {\bibfnamefont {G.-C.}\ \bibnamefont {{Guo}}},\ }\bibfield  {title}
  {\enquote {\bibinfo {title} {{Arbitrary quantum superposition state for
  three-level system using oscillating dark states}},}\ }\href {\doibase
  10.1016/j.optcom.2008.11.078} {\bibfield  {journal} {\bibinfo  {journal}
  {Optics Communications}\ }\textbf {\bibinfo {volume} {282}},\ \bibinfo
  {pages} {1167--1170} (\bibinfo {year} {2009})}\BibitemShut {NoStop}%
\bibitem [{\citenamefont {Randall}\ \emph {et~al.}(2018)\citenamefont
  {Randall}, \citenamefont {Lawrence}, \citenamefont {Webster}, \citenamefont
  {Weidt}, \citenamefont {Vitanov},\ and\ \citenamefont
  {Hensinger}}]{randall2018generation}%
  \BibitemOpen
  \bibfield  {author} {\bibinfo {author} {\bibfnamefont {J.}~\bibnamefont
  {Randall}}, \bibinfo {author} {\bibfnamefont {A.~M.}\ \bibnamefont
  {Lawrence}}, \bibinfo {author} {\bibfnamefont {S.~C.}\ \bibnamefont
  {Webster}}, \bibinfo {author} {\bibfnamefont {S.}~\bibnamefont {Weidt}},
  \bibinfo {author} {\bibfnamefont {N.~V.}\ \bibnamefont {Vitanov}}, \ and\
  \bibinfo {author} {\bibfnamefont {W.~K.}\ \bibnamefont {Hensinger}},\
  }\bibfield  {title} {\enquote {\bibinfo {title} {Generation of high-fidelity
  quantum control methods for multilevel systems},}\ }\href {\doibase
  10.1103/PhysRevA.98.043414} {\bibfield  {journal} {\bibinfo  {journal} {Phys.
  Rev. A}\ }\textbf {\bibinfo {volume} {98}},\ \bibinfo {pages} {043414}
  (\bibinfo {year} {2018})}\BibitemShut {NoStop}%
\bibitem [{\citenamefont {Martens}\ \emph {et~al.}(2013)\citenamefont
  {Martens}, \citenamefont {Longo},\ and\ \citenamefont
  {Busch}}]{martens2013photon}%
  \BibitemOpen
  \bibfield  {author} {\bibinfo {author} {\bibfnamefont {C.}~\bibnamefont
  {Martens}}, \bibinfo {author} {\bibfnamefont {P.}~\bibnamefont {Longo}}, \
  and\ \bibinfo {author} {\bibfnamefont {K.}~\bibnamefont {Busch}},\ }\bibfield
   {title} {\enquote {\bibinfo {title} {Photon transport in one-dimensional
  systems coupled to three-level quantum impurities},}\ }\href {\doibase
  10.1088/1367-2630/15/8/083019} {\bibfield  {journal} {\bibinfo  {journal}
  {New Journal of Physics}\ }\textbf {\bibinfo {volume} {15}},\ \bibinfo
  {pages} {083019} (\bibinfo {year} {2013})}\BibitemShut {NoStop}%
\bibitem [{\citenamefont {Chan}\ \emph {et~al.}(2022)\citenamefont {Chan},
  \citenamefont {Aqua}, \citenamefont {Tiranov}, \citenamefont {Dayan},
  \citenamefont {Lodahl},\ and\ \citenamefont {S\o{}rensen}}]{chan2022quantum}%
  \BibitemOpen
  \bibfield  {author} {\bibinfo {author} {\bibfnamefont {M.~L.}\ \bibnamefont
  {Chan}}, \bibinfo {author} {\bibfnamefont {Z.}~\bibnamefont {Aqua}}, \bibinfo
  {author} {\bibfnamefont {A.}~\bibnamefont {Tiranov}}, \bibinfo {author}
  {\bibfnamefont {B.}~\bibnamefont {Dayan}}, \bibinfo {author} {\bibfnamefont
  {P.}~\bibnamefont {Lodahl}}, \ and\ \bibinfo {author} {\bibfnamefont {A.~S.}\
  \bibnamefont {S\o{}rensen}},\ }\bibfield  {title} {\enquote {\bibinfo {title}
  {Quantum state transfer between a frequency-encoded photonic qubit and a
  quantum-dot spin in a nanophotonic waveguide},}\ }\href {\doibase
  10.1103/PhysRevA.105.062445} {\bibfield  {journal} {\bibinfo  {journal}
  {Phys. Rev. A}\ }\textbf {\bibinfo {volume} {105}},\ \bibinfo {pages}
  {062445} (\bibinfo {year} {2022})}\BibitemShut {NoStop}%
\bibitem [{\citenamefont {Bradford}\ and\ \citenamefont
  {Shen}(2012)}]{bradford2012single}%
  \BibitemOpen
  \bibfield  {author} {\bibinfo {author} {\bibfnamefont {M.}~\bibnamefont
  {Bradford}}\ and\ \bibinfo {author} {\bibfnamefont {J.-T.}\ \bibnamefont
  {Shen}},\ }\bibfield  {title} {\enquote {\bibinfo {title} {Single-photon
  frequency conversion by exploiting quantum interference},}\ }\href {\doibase
  10.1103/PhysRevA.85.043814} {\bibfield  {journal} {\bibinfo  {journal} {Phys.
  Rev. A}\ }\textbf {\bibinfo {volume} {85}},\ \bibinfo {pages} {043814}
  (\bibinfo {year} {2012})}\BibitemShut {NoStop}%
\bibitem [{\citenamefont {{Shen}}\ and\ \citenamefont
  {{Fan}}(2005)}]{shen2005coherent}%
  \BibitemOpen
  \bibfield  {author} {\bibinfo {author} {\bibfnamefont {J.~T.}\ \bibnamefont
  {{Shen}}}\ and\ \bibinfo {author} {\bibfnamefont {S.}~\bibnamefont {{Fan}}},\
  }\bibfield  {title} {\enquote {\bibinfo {title} {{Coherent photon transport
  from spontaneous emission in one-dimensional waveguides}},}\ }\href {\doibase
  10.1364/OL.30.002001} {\bibfield  {journal} {\bibinfo  {journal} {Optics
  Letters}\ }\textbf {\bibinfo {volume} {30}},\ \bibinfo {pages} {2001--2003}
  (\bibinfo {year} {2005})}\BibitemShut {NoStop}%
\bibitem [{\citenamefont {{Witthaut}}\ and\ \citenamefont
  {{S{\o}rensen}}(2010)}]{witthaut2010photon}%
  \BibitemOpen
  \bibfield  {author} {\bibinfo {author} {\bibfnamefont {D.}~\bibnamefont
  {{Witthaut}}}\ and\ \bibinfo {author} {\bibfnamefont {A.~S.}\ \bibnamefont
  {{S{\o}rensen}}},\ }\bibfield  {title} {\enquote {\bibinfo {title} {{Photon
  scattering by a three-level emitter in a one-dimensional waveguide}},}\
  }\href {\doibase 10.1088/1367-2630/12/4/043052} {\bibfield  {journal}
  {\bibinfo  {journal} {New Journal of Physics}\ }\textbf {\bibinfo {volume}
  {12}},\ \bibinfo {eid} {043052} (\bibinfo {year} {2010})}\BibitemShut
  {NoStop}%
\bibitem [{\citenamefont {Shen}\ and\ \citenamefont
  {Fan}(2007)}]{shen2007strongly}%
  \BibitemOpen
  \bibfield  {author} {\bibinfo {author} {\bibfnamefont {J.-T.}\ \bibnamefont
  {Shen}}\ and\ \bibinfo {author} {\bibfnamefont {S.}~\bibnamefont {Fan}},\
  }\bibfield  {title} {\enquote {\bibinfo {title} {Strongly correlated
  multiparticle transport in one dimension through a quantum impurity},}\
  }\href {\doibase 10.1103/PhysRevA.76.062709} {\bibfield  {journal} {\bibinfo
  {journal} {Phys. Rev. A}\ }\textbf {\bibinfo {volume} {76}},\ \bibinfo
  {pages} {062709} (\bibinfo {year} {2007})}\BibitemShut {NoStop}%
\bibitem [{\citenamefont {Schuster}\ \emph {et~al.}(2005)\citenamefont
  {Schuster}, \citenamefont {Wallraff}, \citenamefont {Blais}, \citenamefont
  {Frunzio}, \citenamefont {Huang}, \citenamefont {Majer}, \citenamefont
  {Girvin},\ and\ \citenamefont {Schoelkopf}}]{schuster2005ac}%
  \BibitemOpen
  \bibfield  {author} {\bibinfo {author} {\bibfnamefont {D.~I.}\ \bibnamefont
  {Schuster}}, \bibinfo {author} {\bibfnamefont {A.}~\bibnamefont {Wallraff}},
  \bibinfo {author} {\bibfnamefont {A.}~\bibnamefont {Blais}}, \bibinfo
  {author} {\bibfnamefont {L.}~\bibnamefont {Frunzio}}, \bibinfo {author}
  {\bibfnamefont {R.-S.}\ \bibnamefont {Huang}}, \bibinfo {author}
  {\bibfnamefont {J.}~\bibnamefont {Majer}}, \bibinfo {author} {\bibfnamefont
  {S.~M.}\ \bibnamefont {Girvin}}, \ and\ \bibinfo {author} {\bibfnamefont
  {R.~J.}\ \bibnamefont {Schoelkopf}},\ }\bibfield  {title} {\enquote {\bibinfo
  {title} {ac stark shift and dephasing of a superconducting qubit strongly
  coupled to a cavity field},}\ }\href {\doibase 10.1103/PhysRevLett.94.123602}
  {\bibfield  {journal} {\bibinfo  {journal} {Phys. Rev. Lett.}\ }\textbf
  {\bibinfo {volume} {94}},\ \bibinfo {pages} {123602} (\bibinfo {year}
  {2005})}\BibitemShut {NoStop}%
\bibitem [{\citenamefont {Manga~Rao}\ and\ \citenamefont
  {Hughes}(2007)}]{rao2007single}%
  \BibitemOpen
  \bibfield  {author} {\bibinfo {author} {\bibfnamefont {V.~S.~C.}\
  \bibnamefont {Manga~Rao}}\ and\ \bibinfo {author} {\bibfnamefont
  {S.}~\bibnamefont {Hughes}},\ }\bibfield  {title} {\enquote {\bibinfo {title}
  {Single quantum-dot purcell factor and $\ensuremath{\beta}$ factor in a
  photonic crystal waveguide},}\ }\href {\doibase 10.1103/PhysRevB.75.205437}
  {\bibfield  {journal} {\bibinfo  {journal} {Phys. Rev. B}\ }\textbf {\bibinfo
  {volume} {75}},\ \bibinfo {pages} {205437} (\bibinfo {year}
  {2007})}\BibitemShut {NoStop}%
\bibitem [{\citenamefont {Rephaeli}\ and\ \citenamefont
  {Fan}(2012)}]{rephaeli2012stimulated}%
  \BibitemOpen
  \bibfield  {author} {\bibinfo {author} {\bibfnamefont {E.}~\bibnamefont
  {Rephaeli}}\ and\ \bibinfo {author} {\bibfnamefont {S.}~\bibnamefont {Fan}},\
  }\bibfield  {title} {\enquote {\bibinfo {title} {Stimulated emission from a
  single excited atom in a waveguide},}\ }\href {\doibase
  10.1103/PhysRevLett.108.143602} {\bibfield  {journal} {\bibinfo  {journal}
  {Phys. Rev. Lett.}\ }\textbf {\bibinfo {volume} {108}},\ \bibinfo {pages}
  {143602} (\bibinfo {year} {2012})}\BibitemShut {NoStop}%
\bibitem [{\citenamefont {Shen}\ and\ \citenamefont
  {Fan}(2009)}]{shen2009theory}%
  \BibitemOpen
  \bibfield  {author} {\bibinfo {author} {\bibfnamefont {J.-T.}\ \bibnamefont
  {Shen}}\ and\ \bibinfo {author} {\bibfnamefont {S.}~\bibnamefont {Fan}},\
  }\bibfield  {title} {\enquote {\bibinfo {title} {Theory of single-photon
  transport in a single-mode waveguide. ii. coupling to a whispering-gallery
  resonator containing a two-level atom},}\ }\href {\doibase
  10.1103/PhysRevA.79.023838} {\bibfield  {journal} {\bibinfo  {journal} {Phys.
  Rev. A}\ }\textbf {\bibinfo {volume} {79}},\ \bibinfo {pages} {023838}
  (\bibinfo {year} {2009})}\BibitemShut {NoStop}%
\bibitem [{\citenamefont {{Hempel}}\ \emph {et~al.}(2013)\citenamefont
  {{Hempel}}, \citenamefont {{Lanyon}}, \citenamefont {{Jurcevic}},
  \citenamefont {{Gerritsma}}, \citenamefont {{Blatt}},\ and\ \citenamefont
  {{Roos}}}]{hempel2013entanglement}%
  \BibitemOpen
  \bibfield  {author} {\bibinfo {author} {\bibfnamefont {C.}~\bibnamefont
  {{Hempel}}}, \bibinfo {author} {\bibfnamefont {B.~P.}\ \bibnamefont
  {{Lanyon}}}, \bibinfo {author} {\bibfnamefont {P.}~\bibnamefont
  {{Jurcevic}}}, \bibinfo {author} {\bibfnamefont {R.}~\bibnamefont
  {{Gerritsma}}}, \bibinfo {author} {\bibfnamefont {R.}~\bibnamefont
  {{Blatt}}}, \ and\ \bibinfo {author} {\bibfnamefont {C.~F.}\ \bibnamefont
  {{Roos}}},\ }\bibfield  {title} {\enquote {\bibinfo {title}
  {{Entanglement-enhanced detection of single-photon scattering events}},}\
  }\href {\doibase 10.1038/nphoton.2013.172} {\bibfield  {journal} {\bibinfo
  {journal} {Nature Photonics}\ }\textbf {\bibinfo {volume} {7}},\ \bibinfo
  {pages} {630--633} (\bibinfo {year} {2013})},\ \Eprint
  {http://arxiv.org/abs/1304.3270} {arXiv:1304.3270 [quant-ph]} \BibitemShut
  {NoStop}%
\bibitem [{\citenamefont {Reim}\ \emph {et~al.}(2011)\citenamefont {Reim},
  \citenamefont {Michelberger}, \citenamefont {Lee}, \citenamefont {Nunn},
  \citenamefont {Langford},\ and\ \citenamefont {Walmsley}}]{reim2011single}%
  \BibitemOpen
  \bibfield  {author} {\bibinfo {author} {\bibfnamefont {K.~F.}\ \bibnamefont
  {Reim}}, \bibinfo {author} {\bibfnamefont {P.}~\bibnamefont {Michelberger}},
  \bibinfo {author} {\bibfnamefont {K.~C.}\ \bibnamefont {Lee}}, \bibinfo
  {author} {\bibfnamefont {J.}~\bibnamefont {Nunn}}, \bibinfo {author}
  {\bibfnamefont {N.~K.}\ \bibnamefont {Langford}}, \ and\ \bibinfo {author}
  {\bibfnamefont {I.~A.}\ \bibnamefont {Walmsley}},\ }\bibfield  {title}
  {\enquote {\bibinfo {title} {Single-photon-level quantum memory at room
  temperature},}\ }\href {\doibase 10.1103/PhysRevLett.107.053603} {\bibfield
  {journal} {\bibinfo  {journal} {Phys. Rev. Lett.}\ }\textbf {\bibinfo
  {volume} {107}},\ \bibinfo {pages} {053603} (\bibinfo {year}
  {2011})}\BibitemShut {NoStop}%
\bibitem [{\citenamefont {{Chen}}\ \emph {et~al.}(2008)\citenamefont {{Chen}},
  \citenamefont {{Chen}}, \citenamefont {{Yuan}}, \citenamefont {{Zhao}},
  \citenamefont {{Chuu}}, \citenamefont {{Schmiedmayer}},\ and\ \citenamefont
  {{Pan}}}]{chen2008memory}%
  \BibitemOpen
  \bibfield  {author} {\bibinfo {author} {\bibfnamefont {Y.-A.}\ \bibnamefont
  {{Chen}}}, \bibinfo {author} {\bibfnamefont {S.}~\bibnamefont {{Chen}}},
  \bibinfo {author} {\bibfnamefont {Z.-S.}\ \bibnamefont {{Yuan}}}, \bibinfo
  {author} {\bibfnamefont {B.}~\bibnamefont {{Zhao}}}, \bibinfo {author}
  {\bibfnamefont {C.-S.}\ \bibnamefont {{Chuu}}}, \bibinfo {author}
  {\bibfnamefont {J.}~\bibnamefont {{Schmiedmayer}}}, \ and\ \bibinfo {author}
  {\bibfnamefont {J.-W.}\ \bibnamefont {{Pan}}},\ }\bibfield  {title} {\enquote
  {\bibinfo {title} {{Memory-built-in quantum teleportation with photonic and
  atomic qubits}},}\ }\href {\doibase 10.1038/nphys832} {\bibfield  {journal}
  {\bibinfo  {journal} {Nature Physics}\ }\textbf {\bibinfo {volume} {4}},\
  \bibinfo {pages} {103--107} (\bibinfo {year} {2008})},\ \Eprint
  {http://arxiv.org/abs/0705.1256} {arXiv:0705.1256 [quant-ph]} \BibitemShut
  {NoStop}%
\bibitem [{\citenamefont {{Hwang}}\ \emph {et~al.}(2009)\citenamefont
  {{Hwang}}, \citenamefont {{Pototschnig}}, \citenamefont {{Lettow}},
  \citenamefont {{Zumofen}}, \citenamefont {{Renn}}, \citenamefont
  {{G{\"o}tzinger}},\ and\ \citenamefont {{Sandoghdar}}}]{hwang2009single}%
  \BibitemOpen
  \bibfield  {author} {\bibinfo {author} {\bibfnamefont {J.}~\bibnamefont
  {{Hwang}}}, \bibinfo {author} {\bibfnamefont {M.}~\bibnamefont
  {{Pototschnig}}}, \bibinfo {author} {\bibfnamefont {R.}~\bibnamefont
  {{Lettow}}}, \bibinfo {author} {\bibfnamefont {G.}~\bibnamefont {{Zumofen}}},
  \bibinfo {author} {\bibfnamefont {A.}~\bibnamefont {{Renn}}}, \bibinfo
  {author} {\bibfnamefont {S.}~\bibnamefont {{G{\"o}tzinger}}}, \ and\ \bibinfo
  {author} {\bibfnamefont {V.}~\bibnamefont {{Sandoghdar}}},\ }\bibfield
  {title} {\enquote {\bibinfo {title} {{A single-molecule optical
  transistor}},}\ }\href {\doibase 10.1038/nature08134} {\bibfield  {journal}
  {\bibinfo  {journal} {\nat}\ }\textbf {\bibinfo {volume} {460}},\ \bibinfo
  {pages} {76--80} (\bibinfo {year} {2009})}\BibitemShut {NoStop}%
\bibitem [{\citenamefont {{Vitanov}}\ \emph {et~al.}(1999)\citenamefont
  {{Vitanov}}, \citenamefont {{Suominen}},\ and\ \citenamefont
  {{Shore}}}]{vitanov1999creation}%
  \BibitemOpen
  \bibfield  {author} {\bibinfo {author} {\bibfnamefont {N.~V.}\ \bibnamefont
  {{Vitanov}}}, \bibinfo {author} {\bibfnamefont {K.~A.}\ \bibnamefont
  {{Suominen}}}, \ and\ \bibinfo {author} {\bibfnamefont {B.~W.}\ \bibnamefont
  {{Shore}}},\ }\bibfield  {title} {\enquote {\bibinfo {title} {{Creation of
  coherent atomic superpositions by fractional stimulated Raman adiabatic
  passage}},}\ }\href {\doibase 10.1088/0953-4075/32/18/312} {\bibfield
  {journal} {\bibinfo  {journal} {Journal of Physics B Atomic Molecular
  Physics}\ }\textbf {\bibinfo {volume} {32}},\ \bibinfo {pages} {4535--4546}
  (\bibinfo {year} {1999})}\BibitemShut {NoStop}%
\bibitem [{\citenamefont {Fan}\ and\ \citenamefont
  {Joannopoulos}(2002)}]{fan2002analysis}%
  \BibitemOpen
  \bibfield  {author} {\bibinfo {author} {\bibfnamefont {S.}~\bibnamefont
  {Fan}}\ and\ \bibinfo {author} {\bibfnamefont {J.~D.}\ \bibnamefont
  {Joannopoulos}},\ }\bibfield  {title} {\enquote {\bibinfo {title} {Analysis
  of guided resonances in photonic crystal slabs},}\ }\href {\doibase
  10.1103/physrevb.65.235112} {\bibfield  {journal} {\bibinfo  {journal}
  {Physical Review B}\ }\textbf {\bibinfo {volume} {65}} (\bibinfo {year}
  {2002}),\ 10.1103/physrevb.65.235112}\BibitemShut {NoStop}%
\bibitem [{\citenamefont {Berman}\ and\ \citenamefont
  {O'Connell}(2005)}]{berman2005constraints}%
  \BibitemOpen
  \bibfield  {author} {\bibinfo {author} {\bibfnamefont {P.~R.}\ \bibnamefont
  {Berman}}\ and\ \bibinfo {author} {\bibfnamefont {R.~C.}\ \bibnamefont
  {O'Connell}},\ }\bibfield  {title} {\enquote {\bibinfo {title} {Constraints
  on dephasing widths and shifts in three-level quantum systems},}\ }\href
  {\doibase 10.1103/PhysRevA.71.022501} {\bibfield  {journal} {\bibinfo
  {journal} {Phys. Rev. A}\ }\textbf {\bibinfo {volume} {71}},\ \bibinfo
  {pages} {022501} (\bibinfo {year} {2005})}\BibitemShut {NoStop}%
\bibitem [{\citenamefont {Liu}\ \emph {et~al.}(2020)\citenamefont {Liu},
  \citenamefont {Wu}, \citenamefont {Hu}, \citenamefont {Zhang},\ and\
  \citenamefont {Zhang}}]{liu2020experimental}%
  \BibitemOpen
  \bibfield  {author} {\bibinfo {author} {\bibfnamefont {J.}~\bibnamefont
  {Liu}}, \bibinfo {author} {\bibfnamefont {J.}~\bibnamefont {Wu}}, \bibinfo
  {author} {\bibfnamefont {Y.}~\bibnamefont {Hu}}, \bibinfo {author}
  {\bibfnamefont {Y.}~\bibnamefont {Zhang}}, \ and\ \bibinfo {author}
  {\bibfnamefont {J.}~\bibnamefont {Zhang}},\ }\bibfield  {title} {\enquote
  {\bibinfo {title} {Experimental demonstration of quantum interference
  modulation via precise dephasing control in atoms},}\ }\href
  {https://www.sciencedirect.com/science/article/pii/S0030401820302248}
  {\bibfield  {journal} {\bibinfo  {journal} {Optics Communications}\ }\textbf
  {\bibinfo {volume} {466}},\ \bibinfo {pages} {125655} (\bibinfo {year}
  {2020})}\BibitemShut {NoStop}%
\bibitem [{\citenamefont {Ruostekoski}\ and\ \citenamefont
  {Javanainen}(2017)}]{ruostekoski2017arrays}%
  \BibitemOpen
  \bibfield  {author} {\bibinfo {author} {\bibfnamefont {J.}~\bibnamefont
  {Ruostekoski}}\ and\ \bibinfo {author} {\bibfnamefont {J.}~\bibnamefont
  {Javanainen}},\ }\bibfield  {title} {\enquote {\bibinfo {title} {Arrays of
  strongly coupled atoms in a one-dimensional waveguide},}\ }\href {\doibase
  10.1103/PhysRevA.96.033857} {\bibfield  {journal} {\bibinfo  {journal} {Phys.
  Rev. A}\ }\textbf {\bibinfo {volume} {96}},\ \bibinfo {pages} {033857}
  (\bibinfo {year} {2017})}\BibitemShut {NoStop}%
\end{thebibliography}%


%

\let\addcontentsline\oldaddcontentsline

\newpage

\setcounter{figure}{0}
\setcounter{section}{0}
\setcounter{equation}{0}
\renewcommand{\thefigure}{S\arabic{figure}}
\renewcommand{\thesection}{S\Roman{section}}
\renewcommand{\thesection}{S\arabic{section}}
\renewcommand{\theequation}{S\arabic{equation}}

%

\end{document}